\documentclass[aps,prx,reprint,preprintnumbers,superscriptaddress,nofootinbib,longbibliography,floatfix]{revtex4-2}
\usepackage{rotating}
\usepackage{array}
\usepackage{ragged2e}
\usepackage{caption}
\usepackage{subcaption}
\usepackage[normalem]{ulem}
\usepackage{booktabs}
\usepackage[compat=1.1.0]{tikz-feynman}
\usepackage{makecell}
\usepackage{slashed}
\usepackage{multirow}
\usepackage{units}
\usepackage{xfrac}
\usepackage{mathtools}
\usepackage{empheq}
\usepackage{amssymb}
\usepackage{url}
\usepackage{comment}
\usepackage{physics}
\usepackage{color,soul}
\usepackage{bbm}
\usepackage{adjustbox}
\usepackage[T1]{fontenc}
\usepackage{algorithm}
\usepackage{algpseudocode}

\newcommand{\deffusion}{\textsc{{D${e^+e^-}$}ffusion}}
\newcommand{\gpig}{\textsc{GuineaPig++}}
\newcommand{\gfour}{\textsc{Geant4}}

\newcommand{\pT}{p_\mathrm{T}}
\newcommand{\eps}{\varepsilon}
\newcommand{\bbeta}{\boldsymbol{\beta}}
\newcommand{\bx}{\mathbf{x}}
\newcommand{\bu}{\mathbf{u}}

\newcommand{\ie}{i.e.\ }

\usepackage{hyperref}
\hypersetup{
  colorlinks=true,
  citecolor=blue,
  linkcolor=blue,
  urlcolor=blue
}

\begin{document}

\title{\texorpdfstring{\boldmath \textsc{D$\boldsymbol{e^+e^-}$ffusion}: Capturing the Beam-Beam Physics of $e^+e^-$ Collisions\\ with Diffusion Models}{De+e-fusion: Capturing the Beam-Beam Physics of e+e- Collisions with Diffusion Models}}

\author{Antonio Chahine}
\affiliation{Blackett Laboratory, Imperial College, London, UK}
\author{Mariarosaria D'Alfonso}
\affiliation{Laboratory for Nuclear Science, Massachusetts Institute of Technology, Cambridge, MA, USA}
\author{Jan Eysermans}
\affiliation{Laboratory for Nuclear Science, Massachusetts Institute of Technology, Cambridge, MA, USA}
\author{Emmett Forrestel}
\affiliation{Department of Physics \& Astronomy, Brown University, Providence, RI, USA}
\author{Loukas Gouskos}
\affiliation{Department of Physics \& Astronomy, Brown University, Providence, RI, USA}
\author{Lindsey Gray}
\affiliation{Fermi National Accelerator Laboratory, Batavia, IL, USA}
\author{Katie Kudela}
\affiliation{Laboratory for Nuclear Science, Massachusetts Institute of Technology, Cambridge, MA, USA}
\author{Haoyun Liu}
\affiliation{Blackett Laboratory, Imperial College, London, UK}
\author{Benedikt Maier}
\email{benedikt.maier@cern.ch}
\affiliation{Blackett Laboratory, Imperial College, London, UK}
\author{Dimitrios Ntounis}
\affiliation{Department of Physics, Stanford University, Stanford, CA, USA}
\affiliation{SLAC National Accelerator Laboratory, Menlo Park, CA, USA}
\author{Christoph Paus}
\affiliation{Laboratory for Nuclear Science, Massachusetts Institute of Technology, Cambridge, MA, USA}
\author{Umar Sohail Qureshi}
\email{uqureshi@cern.ch}
\affiliation{Department of Physics, Stanford University, Stanford, CA, USA}
\affiliation{SLAC National Accelerator Laboratory, Menlo Park, CA, USA}
\author{Caterina Vernieri}
\affiliation{SLAC National Accelerator Laboratory, Menlo Park, CA, USA}
\affiliation{Department of Particle Physics and Astrophysics, Stanford University, Stanford, CA, USA}

\begin{abstract}
Beam-induced backgrounds at high-luminosity $e^+e^-$ colliders, such as the FCC-ee, are dominated by incoherent pair creation (IPC), and require computationally expensive simulations with dedicated Monte Carlo (MC) event generators. Reliable detector and machine-detector interface studies necessitate event samples that are several orders of magnitude larger than what is practically attainable with existing MC. To alleviate this bottleneck, we present {\deffusion}, a denoising diffusion probabilistic model that operates as a permutation-equivariant, set-valued surrogate for fast IPC simulation. Trained on a small {\gpig} sample, {\deffusion} faithfully reproduces the marginal and joint kinematic, angular, and positional distributions of all three IPC production processes. In addition, we assess the fidelity at the detector level by propagating both {\gpig} and {\deffusion} events through a {\gfour} simulation of the CLD vertex detector and by training a transformer-based two-sample classifier; the classifier achieves an area under the ROC curve of $0.553 \pm 0.016$. The trained model generates events nearly four orders of magnitude faster than {\gpig}, paving the way for a fast-simulation surrogate for FCC-ee design studies.
\end{abstract}

\maketitle
%\tableofcontents

\section{Introduction and Motivation}

High-fidelity Monte Carlo (MC) simulation samples are central to every stage of an $e^+e^-$ collider development program. At the FCC-ee~\cite{FCCee_FSR}, beam-beam interactions in the strong electromagnetic fields of dense relativistic bunches generate substantial beam-induced backgrounds (BIB) that drive the design of the innermost detector layers, the machine-detector interface, and the beam-pipe geometry. The dominant component of these luminosity-driven backgrounds is incoherent pair creation (IPC), in which the equivalent photon fields of the two bunches~\cite{Yokoya:1991} produce $e^+e^-$ pairs through three QED subprocesses: the real-photon Breit--Wheeler reaction $\gamma\gamma\to e^+e^-$~\cite{Breit:1934}, the mixed real-virtual Bethe--Heitler reaction $\gamma\gamma^\ast\to e^+e^-$~\cite{Bethe:1934}, and the doubly-virtual Landau--Lifshitz reaction $\gamma^\ast\gamma^\ast\to e^+e^-$~\cite{Landau:1934}. The Landau--Lifshitz cross section, which scales as $\sigma\propto \ln^3 s$, is the dominant source of forward incoherent pairs at FCC-ee and produces the so-called \emph{pair envelope} that constrains the inner-tracker geometry.

The de facto standard generator used by the FCC-ee Collaboration for these processes is {\gpig}, a strong-field tracking particle-in-cell code that integrates the classical equations of motion of macroparticles in self-consistently computed beam fields and that samples QED final states. We refer the interested reader to Refs.~\cite{Eysermans_GUINEA-PIG_fork, Rimbault:2007, Schulte:1998}. The {\gpig} generator is, however, computationally expensive, requiring several CPU-hours to generate a single bunch crossing, and thus the per-event cost is dominated by the strong-field calculation of IPCs rather than by the QED matrix element or detector simulation. The samples required for a comprehensive detector-occupancy, tracking, or radiation-damage study at FCC-ee are at the level of $\mathcal{O}(10^5)$ bunch crossings, beyond what is computationally feasible during the design phase.

This is reminiscent of computational regimes in which generative machine learning has, in recent years, become a competitive surrogate for full-simulation chains in collider physics. For instance, the \textsc{Parnassus} program~\cite{Dreyer:2024parnassus} has demonstrated that continuous normalizing flows and diffusion models conditioned on a truth point cloud can replace the entire CMS particle-flow simulation-reconstruction pipeline at the jet level. Diffusion-based models, originally developed for natural images~\cite{Sohl2015,Ho2020,Song2021}, share with normalizing flows the advantage of stable likelihood-based training, but also support efficient sampling of high-dimensional, strongly correlated point clouds. In particle physics, diffusion models have already been applied to calorimeter showers~\cite{Mikuni:2022,Buhmann:2023}, jet generation~\cite{Leigh:2023,Mikuni:2023} and pile-up~\cite{KasielukJets:2024}.

In this work, we introduce {\deffusion}, a diffusion model fast simulation surrogate of the IPC component of FCC-ee beam backgrounds. The model jointly diffuses the continuous kinematic and spatial state of each particle together with its discrete charge label, while the underlying QED subprocess (Breit--Wheeler, Bethe--Heitler, or Landau--Lifshitz) is supplied as a time-independent conditioning input. This conditioning, combined with an inverse-frequency reweighting of the denoising loss, allows the model to reproduce the distinct kinematic and positional structure of each IPC channel, which a process-agnostic model tends to wash out.

We evaluate the fidelity of the surrogate using three complementary strategies: (i) we evaluate the non-closure of one-dimensional marginals of all generated kinematic, angular, and spatial features; (ii) we compare per-process longitudinal distributions; and (iii)~we perform a transformer-based two-sample classifier test, in which both {\gpig} and generated events are passed through a {\gfour}-based~\cite{Geant4:2003} full simulation of the CLD vertex detector~\cite{CLD:2019} and the classifier is asked to discriminate between the resulting hit collections.

The remainder of the paper is organized as follows. Section~\ref{sec:bib} reviews the physics of beam-induced backgrounds at $e^+e^-$ colliders. Section~\ref{sec:dataset} describes the training dataset, the input representation, and the per-particle feature transformations. Section~\ref{sec:method} develops the diffusion model architecture, including the joint continuous-discrete kernel, the process conditioning, the denoiser architecture, and the training and sampling algorithms. Section~\ref{sec:results} reports the results, including the two-sample classifier test and the runtime comparison. We conclude with a short discussion in Section~\ref{sec:discussion}.

\section{\texorpdfstring{\boldmath Beam-Induced Backgrounds at $e^+e^-$ Colliders}%
{Beam-Induced Backgrounds at e+e- Colliders}}
\label{sec:bib}

\subsection{Beamstrahlung Radiation and Incoherent Pair Creation}

When two relativistic bunches of charge $Ne$ and Lorentz factor $\gamma_b$ approach each other, the time-dependent classical electromagnetic field of one bunch can be Lorentz-transformed into a flux of nearly real ``equivalent'' photons in the rest frame of a test particle in the opposing bunch~\cite{Yokoya:1991}. In addition, beamstrahlung, \ie the synchrotron radiation emitted by individual leptons in the collective field, produces a flux of real photons of energy $\omega\sim \gamma_b\Upsilon m_e$, where the beamstrahlung parameter is given as:
\begin{equation}
\Upsilon = \frac{2}{3}\frac{\hbar\omega_c}{m_e c^2 \gamma_b}
= \frac{5r_e^2\gamma_bN}{6\alpha\sigma_z(\sigma_x+\sigma_y)},
\label{eq:upsilon}
\end{equation}
with $r_e$ the classical electron radius, $\alpha$ the fine-structure constant, and $\sigma_x$, $\sigma_y$, $\sigma_z$ the bunch sizes. At the FCC-ee $Z$ pole one has $\Upsilon\sim 10^{-4}$, well in the classical beamstrahlung regime.

\begin{figure*}
  \centering
  % --- (a) Breit--Wheeler ---
  \begin{subfigure}[t]{0.31\linewidth}
    \centering
    \begin{tikzpicture}[scale=1.75, transform shape]
      \begin{feynman}
        \vertex (v);
        \vertex [above left=1.2cm of v]  (g1) {\(\gamma\)};
        \vertex [below left=1.2cm of v]  (g2) {\(\gamma\)};
        \vertex [above right=1.2cm of v] (e1) {\(e^-\)};
        \vertex [below right=1.2cm of v] (e2) {\(e^+\)};
        \diagram*{
          (g1) -- [photon] (v),
          (g2) -- [photon] (v),
          (v)  -- [fermion] (e1),
          (e2) -- [fermion] (v),
        };
      \end{feynman}
      \fill[white] (v) circle (0.175cm);
      \filldraw[pattern=north east lines, draw=black, line width=0.5pt] (v) circle (0.175cm);
    \end{tikzpicture}
    \caption{Breit--Wheeler}
  \end{subfigure}
  \hfill
  % --- (b) Bethe--Heitler ---
  \begin{subfigure}[t]{0.31\linewidth}
    \centering
    \begin{tikzpicture}[scale=1.75, transform shape]
      \begin{feynman}
        \vertex (v);
        \vertex [above left=1.2cm of v]  (g1) {\(\gamma\)};
        \vertex [below left=1.2cm of v]  (g2) {\(\gamma^*\)};
        \vertex [above right=1.2cm of v] (e1) {\(e^-\)};
        \vertex [below right=1.2cm of v] (e2) {\(e^+\)};
        \diagram*{
          (g1) -- [photon] (v),
          (g2) -- [photon] (v),
          (v)  -- [fermion] (e1),
          (e2) -- [fermion] (v),
        };
      \end{feynman}
      \fill[white] (v) circle (0.175cm);
      \filldraw[pattern=north east lines, draw=black, line width=0.5pt] (v) circle (0.175cm);
    \end{tikzpicture}
    \caption{Bethe--Heitler}
  \end{subfigure}
  \hfill
  % --- (c) Landau--Lifshitz ---
  \begin{subfigure}[t]{0.31\linewidth}
    \centering
    \begin{tikzpicture}[scale=1.75, transform shape]
      \begin{feynman}
        \vertex (v);
        \vertex [above left=1.2cm of v]  (g1) {\(\gamma^*\)};
        \vertex [below left=1.2cm of v]  (g2) {\(\gamma^*\)};
        \vertex [above right=1.2cm of v] (e1) {\(e^-\)};
        \vertex [below right=1.2cm of v] (e2) {\(e^+\)};
        \diagram*{
          (g1) -- [photon] (v),
          (g2) -- [photon] (v),
          (v)  -- [fermion] (e1),
          (e2) -- [fermion] (v),
        };
      \end{feynman}
      \fill[white] (v) circle (0.175cm);
      \filldraw[pattern=north east lines, draw=black, line width=0.5pt] (v) circle (0.175cm);
    \end{tikzpicture}
    \caption{Landau--Lifshitz}
  \end{subfigure}
  \caption{\justifying Representative Feynman diagrams for incoherent $e^+e^-$ pair production from beam-beam interactions. Two photons (real beamstrahlung or virtual Coulomb-field photons) interact to produce an $e^+e^-$ pair.}
  \label{fig:feynman-pairs}
\end{figure*}
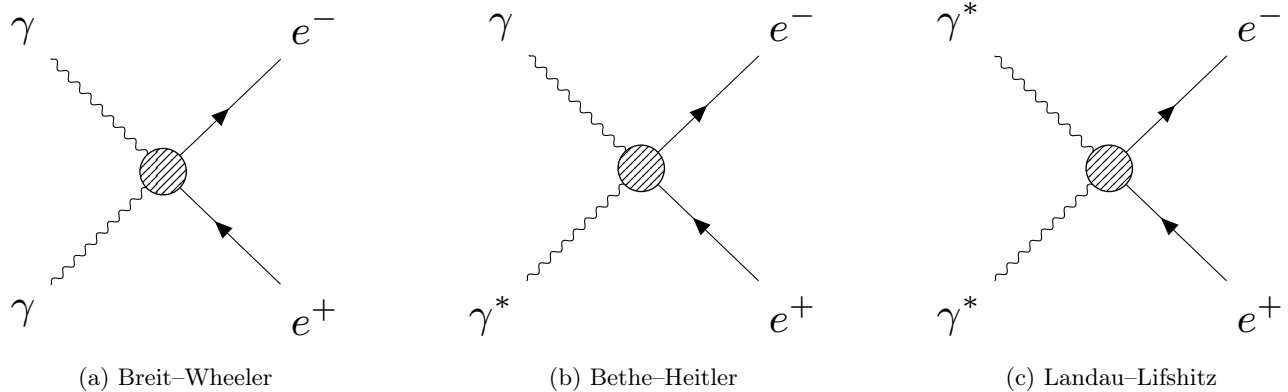

The colliding photon fluxes drive the three IPC subprocesses as shown in Fig.~\ref{fig:feynman-pairs}:
\begin{align}
\gamma\gamma     &\to e^+ e^- & &\text{(Breit--Wheeler, BW)},\\
\gamma\gamma^*   &\to e^+ e^- & &\text{(Bethe--Heitler, BH)},\\
\gamma^*\gamma^* &\to e^+ e^- & &\text{(Landau--Lifshitz, LL)},
\end{align}
where real (on-shell) photons are represented by $\gamma$ and virtual (off-shell) photons by $\gamma^*$. At leading order in $\alpha$ all three are described by the usual $\gamma\gamma\to e^+e^-$ matrix element with appropriate equivalent-photon spectra for the virtual photons. Their relative cross sections are controlled by the real/virtual ratio of the photon spectra and by the available phase space. For typical FCC-ee $Z$-pole parameters at $\sqrt{s}=91.2$\,GeV, the cross sections for the different subprocesses are about $\sigma_\mathrm{BW}:\sigma_\mathrm{BH}:\sigma_\mathrm{LL}\approx 1:14:85$.

The three subprocesses populate distinct regions of the seven-dimensional single-particle phase space $(E,\bbeta,\mathbf{r})$. A discriminating feature between these subprocesses is the longitudinal position coordinate $z$. The longitudinal production point reflects the photon content of each subprocess: BW (real-real) pairs are produced where the two beamstrahlung photons interact and are therefore the most central; LL (virtual-virtual) pairs are produced at the bunches themselves and are the most forward; BH (real-virtual) pairs are intermediate, with a moderately broadened $z$ profile. A model that does not separately resolve the three channels may average the modalities and only learn the marginals over them; this is the motivation for the process conditioning of Section~\ref{sec:method}.

\subsection{Detector Design Considerations}

At FCC-ee the IPC flux dominates the occupancy of the innermost vertex detector layers. Two handles control the impact of this flux on the detector design:
\begin{itemize}
\item \emph{Transverse-momentum spectrum.} The pair $\pT$ is $\mathcal{O}(\mathrm{MeV})$, so most particles spiral in the solenoid and never reach the active layers. The pair envelope, defined as the radius beyond which a fraction $1-\eta$ of particles are curled back, fixes the minimum allowed beam-pipe radius.
\item \emph{Angular spectrum.} The pair system is strongly boosted along the beam axis; the forward calorimeter and luminosity monitors absorb a disproportionate flux.
% \item \emph{Longitudinal position distribution.} Particles produced near $z\approx \pm\sigma_z/2$ traverse a longer path inside the detector and contribute disproportionately to the radiation-damage budget of the forward sensors.
\end{itemize}
A useful surrogate must therefore reproduce the correct joint distributions in $(\pT,\eta,z)$ including non-trivial correlations and not just the one-dimensional marginals. We will see later in Section~\ref{sec:results}, where we demonstrate that this is achieved.

\section{Samples and Simulation}
\label{sec:dataset}

\subsection{Training Sample}

We train on $N_\mathrm{ev}=6\times10^4$ bunch crossings produced with {\gpig} v2.0 using the FCC-ee $Z$-pole machine parameters of the GHC lattice version 25.1 (Feasibility Study Report~\cite{FCCee_FSR}); the relevant inputs are summarized in Table~\ref{tab:gp_params}. Each crossing yields a variable number $K_n$ of $e^\pm$ particles with median $\langle K\rangle\approx 5\times10^2$ and tail extending past $K=10^3$. Samples were produced using an updated version \cite{Eysermans_GUINEA-PIG_fork} of {\gpig} where the following modifications were made to the {\gpig} beam-beam simulation code to support FCC-ee studies. First, a longitudinal solenoid magnetic field $B_z$, the detector magnetic field at the interaction point, was incorporated into the pair-particle tracking. Previously, only the transverse beam-beam electromagnetic fields $(E_x, E_y, B_x, B_y)$ were applied to secondary pairs. The rotation of a particle's velocity vector in the combined field is now computed using the three-dimensional Rodrigues rotation formula. Second, a geometric beam-pipe absorption condition was introduced. Each pair particle is propagated only while it remains inside the cylindrical beam-pipe volume of inner radius $r_\mathrm{bp}$ which is set to $r_\mathrm{bp} = 10$\,mm for this study. At each tracking step, the code solves for the parametric intersection of the particle trajectory with all seven bounding surfaces (the cylinder, the four transverse and two longitudinal grid planes) using a quadratic formula, clamps the step to the earliest crossing, and flags the particle as absorbed. Subsequent field evaluations and position advances are not performed for absorbed particles. Finally, the transverse multi-grid hierarchy was extended from a maximum of seven levels to eight, with a new intermediate grid of lateral size $3\sigma_x \times 3\sigma_x$
inserted between the previous $2\sigma$ and $6\sigma$ outermost levels. This extension was motivated by the significantly larger transverse beam sizes at the FCC-ee $Z$-pole compared to CLIC, for which the original hierarchy was designed. As a result, the deflected pairs are tracked accurately across a wider phase-space volume.

The dataset is randomly split 90/10 into training and validation subsets.

\begin{table}[t]
\centering
\caption{\justifying Inputs to {\gpig} at the $Z$ pole for the GHC (FSR) lattice. A crossing angle of 15\,mrad is used throughout.}
\renewcommand{\arraystretch}{1.25}
\setlength{\tabcolsep}{10pt}

\begin{tabular*}{\columnwidth}{@{\extracolsep{\fill}}l c c}
\toprule
\textbf{Parameter} & \textbf{Units} & \textbf{Value} \\
\midrule
$\sigma_x$            & nm        & 8837 \\
$\sigma_y$            & nm        & 38.3 \\
$\sigma_z$            & $\mu$m    & 15200 \\
$\beta_x^\ast$        & mm        & 110 \\
$\beta_y^\ast$        & mm        & 0.700 \\
Bunch Intensity       & $10^{10}$ & 21.8 \\
Crossing angle        & mrad      & 15 \\
\bottomrule
\end{tabular*}

\label{tab:gp_params}
\end{table}
\begin{figure*}[t]
\centering
\includegraphics[width=0.99\linewidth]{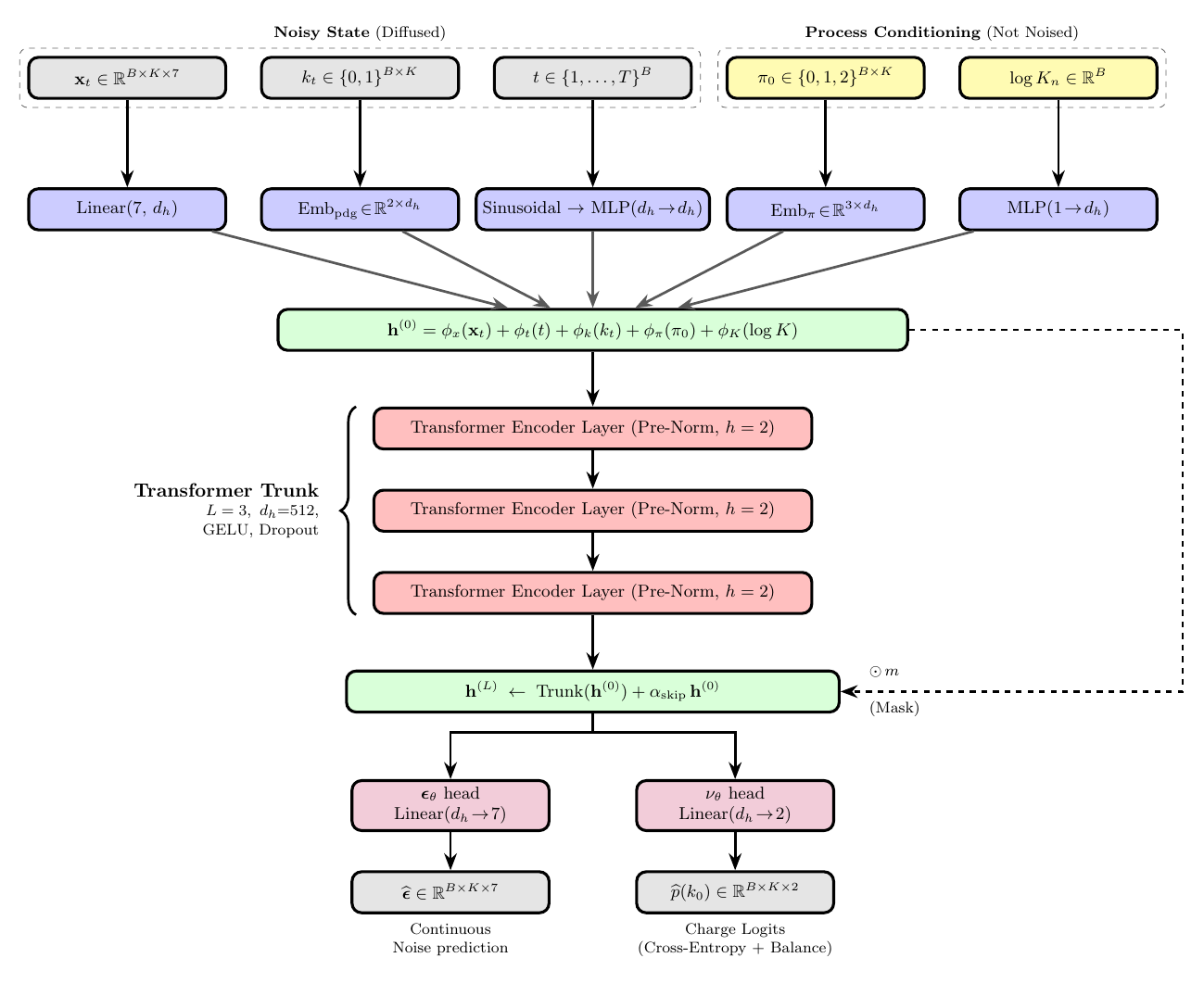}
\caption{\justifying Architecture of the {\deffusion} denoiser. A pre-norm transformer encoder ($L=3$, $h=2$, $d_h=512$) is shared between the noise-prediction head $\boldsymbol{\epsilon}_\theta$ (continuous, output $\mathbb{R}^7$) and the charge head $\nu_\theta$ (discrete, output $\mathbb{R}^2$). The diffusion step $t$, the per-particle charge embedding, the per-particle QED-process embedding and the global log-multiplicity embedding are projected to dimension $d_h$ and additively fused with the projected continuous state into the input embedding $\mathbf{h}^{(0)}$, which is then passed through the trunk together with a residual skip of coefficient $\alpha_\mathrm{skip}=0.2$. The process and multiplicity labels are clean, time-independent conditioning inputs and are never noised; the continuous state $\mathbf{x}_t$ and the charge label $k_t$ are.}
\label{fig:arch}
\end{figure*}

\subsection{Particle-Level Features}

Each bunch crossing $n$ is represented as a set or ``particle cloud'' of $K_n$ particles as:
\begin{equation}
\mathcal{E}_n = \bigl\{(E_i,\bbeta_i,\mathbf{r}_i,c_i,p_i)\bigr\}_{i=1}^{K_n},
\label{eq:event}
\end{equation}
where $E_i\in\mathbb{R}^+$ is the energy in GeV, $\bbeta_i\in B^3$ is the velocity vector in units of $c$ (constrained to $\|\bbeta_i\|<1$), $\mathbf{r}_i=(x_i,y_i,z_i)\in\mathbb{R}^3$ is the position in nm, $c_i\in\{e^-,e^+\}$ is the charge category, and $p_i\in\{\mathrm{BW},\mathrm{BH},\mathrm{LL}\}$ is the QED subprocess label. No ordering is imposed on the particle elements of $\mathcal{E}_n$, so that it is invariant under permutation of $i$. To map this raw representation onto a domain that is well-suited to a Gaussian forward diffusion process we apply three particle-wise transformations.

\paragraph{Log-energy.} The IPC energy spectrum is steeply falling and thus has a long-tailed distribution spanning several orders of magnitude. To accommodate this, we apply a logarithmic transformation as follows:
\begin{equation}
\widetilde{E}_i = \log E_i
\label{eq:logE}
\end{equation}
which compresses the dynamic range and makes the marginal approximately Gaussian. During sampling, one simply recovers the original energy using the inverse $E_i = \exp \widetilde{E}_i$.

\paragraph{Velocity unsquashing.} Direct diffusion of $\bbeta$ would frequently push particles outside the unit ball $\|\bbeta\|<1$ and produce unphysical superluminal samples. We map the unit ball to $\mathbb{R}^3$ via the inverse hyperbolic tangent applied along the magnitude direction, with the direction $\widehat{\bbeta}$ preserved:
\begin{equation}
\bu_i = \mathrm{atanh}\left(\frac{\|\bbeta_i\|}{1-\eps}\right)\widehat{\bbeta}_i,
\label{eq:unsquash}
\end{equation}
where $\eps=10^{-6}$ is chosen so we are always a small offset away from the lightlike boundary. The inverse map, used at decoding is then:
\begin{equation}
\bbeta_i = (1-\eps)\tanh(\|\bu_i\|)\widehat{\bu}_i,
\label{eq:squash}
\end{equation}
which is bijective on $\mathbb{R}^3\setminus\{0\}\to B^3\setminus\{0\}$. Because the magnitude $\|\bu_i\|$ is unbounded above, training the diffusion model on $\bu_i$ is well-posed and never produces $\|\bbeta\|\geq 1$ at sampling time.

\paragraph{Charge encoding.} The $\mathrm{PDG}$ ID $c_i\in\{+11,-11\}$ is mapped to a binary class index $k_i\in\{0,1\}$ ($e^-\mapsto 0$, $e^+\mapsto 1$).

\paragraph{Process encoding.} The subprocess label $p_i\in\{\mathrm{BW},\mathrm{BH},\mathrm{LL}\}$ is mapped to $\pi_i\in\{0,1,2\}$. This label is used only as conditioning; it is never noised (Section~\ref{sec:method:conditioning}).

After the transformations, the per-particle continuous state is the seven-dimensional vector:
\begin{equation}
\bx_i^{(0)} = (\widetilde{E}_i, u_{i,x}, u_{i,y}, u_{i,z}, x_i, y_i, z_i)\in\mathbb{R}^7.
\label{eq:x0}
\end{equation}
A per-feature affine standardization $\bx_i^{(0)} \mapsto (\bx_i^{(0)}-\boldsymbol{\mu})/\boldsymbol{\sigma}$ is applied using the empirical mean $\boldsymbol{\mu}$ and standard deviation $\boldsymbol{\sigma}$ computed once over the training set. The resulting standardized features serve as the input to the model. Because the standardization is bijective and known, all subsequent results below are reported in physical units after the inverse transformation.

\subsection{Padding, Multiplicity and Event Composition}

Diffusion is performed on a fixed-length sequence of $K_{\max}=1300$ particle slots per event. For an event with $K_n\le K_{\max}$ particles we set:
\begin{equation}
\bx^{(0)}_{n,i}=\begin{cases}\overline{\bx}_{n,i}&i<K_n,\\ \mathbf{0}&i\ge K_n,\end{cases}\qquad m_{n,i}=\mathbbm{1}[i<K_n],
\end{equation}
where $\overline\bx_{n,i}$ is the standardized feature vector and $m_{n,i}\in\{0,1\}$ is a Boolean mask. Loss terms, the transformer self-attention and reverse-process updates are all masked by $m$. We do not learn the multiplicity distribution $p(K)$. Instead we cache, using the training dataset:
\begin{itemize}
\item The empirical multiplicity histogram $\{K_n\}_{n=1}^{N_\mathrm{ev}}$, and
\item The per-event process composition $\boldsymbol{f}_n=(f_n^\mathrm{BW},f_n^\mathrm{BH},f_n^\mathrm{LL})$ defined by the empirical fraction of each process within event $n$.
\end{itemize}
At sampling time (Section~\ref{sec:sampling}) a real event $n^\star$ is drawn uniformly at random and supplies both $K_{n^\star}$ and $\boldsymbol{f}_{n^\star}$ to the model. This non-parametric handling of multiplicity and composition is exact by construction up to the resolution of the empirical sample.

\section{Diffusion Model Architecture}
\label{sec:method}

\subsection{Overview}

{\deffusion} is a denoising diffusion probabilistic model (DDPM)~\cite{Ho2020} extended in two ways. First, the per-particle data $(\bx_i,k_i)$ is a heterogeneous tuple of a continuous vector and a categorical label, and we diffuse the two with separate but jointly-denoised kernels. Second, the QED subprocess $\pi_i$ enters the model only as a deterministic conditioning input that is never noised. The forward (noising) process and the reverse (denoising, generative) process are detailed below.

\subsection{Forward Process}

\subsubsection{Continuous Block}

Let $T=10^3$ and let $\{\overline\alpha_t\}_{t=0}^T$ be the cumulative-product schedule of the cosine $\beta$-schedule from Ref.~\cite{Nichol:2021}, as:
\begin{equation}
\overline\alpha_t = \frac{f(t)}{f(0)},\qquad f(t)=\cos^2\left(\frac{t/T+s}{1+s}\frac{\pi}{2}\right),
\label{eq:cosine}
\end{equation}
with $s$ a small offset (we use $s=2.27\times 10^{-3}$). The associated per-step noise schedule is $\beta_t=1-\overline\alpha_t/\overline\alpha_{t-1}$, clamped to $[0,0.999]$. The forward kernel acting on the continuous block is the standard variance-preserving Gaussian diffusion, given as:
\begin{equation}
q(\bx_t|\bx_0) = \mathcal{N}\left(\bx_t; \sqrt{\overline\alpha_t}\bx_0, (1-\overline\alpha_t)\mathbf{I}\right),
\label{eq:fwd_cont}
\end{equation}
which admits the reparameterized sample:
\begin{equation}
\bx_t = \sqrt{\overline\alpha_t}\bx_0 + \sqrt{1-\overline\alpha_t}\boldsymbol{\eps},\qquad \boldsymbol{\eps}\sim\mathcal{N}(0,\mathbf{I}).
\label{eq:reparam}
\end{equation}
For $t=T$ the marginal $q(\bx_\mathrm{T})\to \mathcal{N}(0,\mathbf{I})$ serves as the prior used at sampling. Equation~\eqref{eq:reparam} is applied independently to each particle slot, with the masked rows kept at zero.

\subsubsection{Discrete Block: Uniform-Flip Kernel for Charge}

The charge label $k_i\in\{0,1\}$ is diffused with a discrete uniform-flip kernel of the D3PM family~\cite{Hoogeboom:2021argmax,Austin:2021d3pm}. With $\gamma_t\in[0,1]$ the per-step flip probability,
\begin{equation}
q(k_t | k_{t-1}) =
(1-\gamma_t)\delta_{k_t,k_{t-1}} + \gamma_t\frac{1}{C},
\end{equation}
where $C=2$ is the number of categories. The corresponding stationary distribution is uniform on $\{0,\ldots,C-1\}$, mirroring the role of $\mathcal{N}(0,\mathbf{I})$ in the continuous block. We use a linear schedule $\gamma_t = \gamma_0 + (\gamma_1-\gamma_0)t/T$ with $\gamma_0=5.2\times10^{-5}$ and $\gamma_1=0.18$. The marginal at time $t$ is the closed-form:
\begin{equation}
q(k_t | k_0) = (1-\overline\gamma_t)\delta_{k_t,k_0} + \overline\gamma_t\frac{1}{C},
\end{equation}
with $\overline\gamma_t$ the cumulative mixture weight; in our regime $\overline\gamma_\mathrm{T}\approx 1-\prod_{t=1}^{T}(1-\gamma_t)\to 1-e^{-\int\gamma}$ rapidly approaches uniform.

\subsubsection{Process Labels}
\label{sec:method:conditioning}

The subprocess $\pi_i\in\{0,1,2\}$ is held fixed:
\begin{equation}
q(\pi_t | \pi_0) = \delta_{\pi_t,\pi_0},\qquad t=0,\ldots,T.
\end{equation}
This is the central modeling choice of the present work and we refer to it as ``process conditioning''. At every reverse step the denoiser is told the ground-truth process of every particle. We motivate this choice further in Section~\ref{sec:results}, where we observe that it makes the longitudinal $z$ coordinate distribution reproducible.

\subsection{Reverse Process}

The Markovian reverse process is parameterized by a single neural network $\mu_\theta$ (and an auxiliary classifier head $\nu_\theta$) operating on the joint state. For the continuous block, denoising score matching~\cite{Vincent:2011,Ho2020} reduces the variational lower bound (VLB)~\cite{Sohl2015,Kingma:2021vdm} to predicting the noise $\boldsymbol{\eps}$ in Eq.~\eqref{eq:reparam}:
\begin{equation}
\mathcal{L}_\mathrm{DSM}(\theta) = \mathbb{E}_{t,\bx_0,\boldsymbol{\eps},c}\left[\bigl\|\boldsymbol{\eps}-\boldsymbol{\eps}_\theta(\bx_t,t,k_t,\pi_0)\bigr\|^2\right],
\label{eq:Ldsm}
\end{equation}
where $t\sim\mathrm{Unif}\{1,\ldots,T\}$, the conditioning $c=(k_t,\pi_0)$ is the joint discrete state, and the loss is summed over particles and masked. Once $\boldsymbol{\eps}_\theta$ is trained the reverse continuous transition takes the standard DDPM form as:
\begin{align}
\label{eq:reverse_cont}
\boldsymbol{\mu}_\theta(\bx_t,t)
&= \frac{1}{\sqrt{\alpha_t}}\left(\bx_t-\frac{\beta_t}{\sqrt{1-\overline\alpha_t}}\boldsymbol{\eps}_\theta\right),\nonumber\\
\sigma_t^2 &= \widetilde\beta_t = \frac{(1-\overline\alpha_{t-1})\beta_t}{1-\overline\alpha_t},\\
\bx_{t-1}|\bx_t &\sim \mathcal{N}\bigl(\boldsymbol{\mu}_\theta(\bx_t,t), \sigma_t^2\mathbf{I}\bigr).\nonumber
\end{align}
For the discrete charge block the reverse kernel is a categorical distribution parameterized by the network logits $\nu_\theta(\cdot)\in\mathbb{R}^C$. We do not derive it from a closed-form posterior; instead we use the simplified ``$\bx_0$-prediction'' form of D3PM~\cite{Austin:2021d3pm}, in which $\nu_\theta$ predicts $p_\theta(k_0|\bx_t,k_t,\pi_0)$ at every step and the next $k_{t-1}$ is sampled from this categorical:
\begin{equation}
k_{t-1}|(\bx_t,k_t,\pi_0)\sim\mathrm{Cat}\bigl(\sigma\left(\nu_\theta(\bx_t,t,k_t,\pi_0)\right)\bigr),
\label{eq:reverse_disc}
\end{equation}
with $\sigma(\cdot)$ denoting the softmax activation function. Because the auxiliary loss below is the cross-entropy of this categorical against the ground-truth $k_0$, the discrete kernel is consistent with the marginal posterior at $t=0$.

\subsection{Auxiliary Objectives and Class Reweighting}
\label{sec:method:aux}

The denoising score-matching objective of Eq.~\eqref{eq:Ldsm} trains only the continuous head $\boldsymbol{\eps}_\theta$ and treats every particle as exchangeable. Neither property is appropriate for the IPC setting. First, the charge label is a generated quantity that must be supervised independently, since charge conservation imposes a hard constraint on the global $e^+/e^-$ counts of each event, and the three QED subprocesses populate the dataset with vastly unequal frequencies. We therefore augment the loss with two auxiliary terms and reweight the score-matching contribution.

The categorical head $\nu_\theta$ is trained on the per-particle cross-entropy of its softmax against the ground-truth charge $k_0$,
\begin{equation}
\mathcal{L}_\mathrm{pdg}(\theta) = -\mathbb{E}_{t,\bx_0,k_0,\pi_0}\left[\log p_\theta(k_0\,|\,\bx_t,k_t,\pi_0)\right].
\label{eq:Lpdg}
\end{equation}
This term plays the role of the discrete-state ELBO in the D3PM framework~\cite{Austin:2021d3pm} under the simplified $\bx_0$-parameterization: it is the data-likelihood contribution at $t=0$, and it is what fixes the charge head to predict the ground-truth posterior at every reverse step. Without it the categorical kernel of Eq.~\eqref{eq:reverse_disc} is unconstrained and the model is free to converge to any charge marginal that is consistent with the continuous likelihood.

Charge conservation provides an additional, physics-motivated constraint. The IPC final state consists exclusively of $e^+e^-$ pairs, so every event must be globally neutral. We enforce this prior softly through a regularizer on the expected per-event net charge. With $p^{(-)}_{n,i}=\sigma(\nu_\theta)_{n,i,0}$ and $p^{(+)}_{n,i}=\sigma(\nu_\theta)_{n,i,1}$, we add the loss:
\begin{equation}
\mathcal{L}_\mathrm{bal}(\theta) = \mathbb{E}_n\left[\Bigl(\sum_i m_{n,i}\bigl(p^{(-)}_{n,i}-p^{(+)}_{n,i}\bigr)\Bigr)^{2}\right].
\label{eq:Lbal}
\end{equation}
The third correction targets the strong class imbalance between the three IPC subprocesses, which populate the training set with relative frequencies $f_\mathrm{BW}\approx 0.01$, $f_\mathrm{BH}\approx 0.14$, and $f_\mathrm{LL}\approx 0.85$. Under the uniform per-particle expectation of Eq.~\eqref{eq:Ldsm} the gradient is dominated by Landau--Lifshitz (LL) particles by nearly two orders of magnitude, and the marginal continuous distributions of the rare channels are systematically biased toward the LL distributions. We mitigate this by attaching a soft inverse-frequency weight $w_p\in\mathbb{R}^+$ to each training particle of subprocess $p$, given as:
\begin{equation}
w_p \propto N_p^{-\alpha_w},\qquad \alpha_w=\tfrac{1}{2},
\end{equation}
where $N_p$ is the global per-process particle count and $\alpha_w=1/2$ interpolates between uniform weighting ($\alpha_w=0$) and full inverse-frequency weighting ($\alpha_w=1$); the square-root choice keeps the gradient variance of the rarest class manageable while still upweighting it by roughly a factor of 7. The raw weights are normalized so that their training-set mean is unity and then clipped to $[w_{\min},w_{\max}]=[0.2,5.0]$ to bound the worst-case influence of any single particle, giving:
\begin{equation}
w_p = \mathrm{clip}\left(\frac{N_p^{-1/2}}{\langle N^{-1/2}\rangle},\,w_{\min},\,w_{\max}\right).
\end{equation}
The reweighted denoising loss is the masked, weighted average over the batch as follows:
\begin{equation}
\widetilde{\mathcal{L}}_\mathrm{DSM}(\theta) = \frac{\mathbb{E}_{n,t,\boldsymbol{\eps}}\sum_{i} m_{n,i}\,w_{\pi_{n,i}}\bigl\|\boldsymbol{\eps}-\boldsymbol{\eps}_\theta\bigr\|^2_{n,i}}{\mathbb{E}_n\sum_{i} m_{n,i}\,w_{\pi_{n,i}}}.
\label{eq:Ltilde}
\end{equation}
The reweighting affects the empirical loss but not the underlying target distribution and the model is still trained against an unbiased estimator of the joint particle density, and only the effective sample size of the gradient is modified. In practice this reweighting allows the BW and BH channels to be recovered with approximately the same fidelity as LL despite their much smaller representation in the training data.

The total per-batch objective is then the linear combination given by:
\begin{equation}
\mathcal{L}(\theta) = \widetilde{\mathcal{L}}_\mathrm{DSM}(\theta) + \lambda_\mathrm{pdg}\,\mathcal{L}_\mathrm{pdg}(\theta) + \lambda_\mathrm{bal}\,\mathcal{L}_\mathrm{bal}(\theta),
\label{eq:Ltot}
\end{equation}
with $\lambda_\mathrm{pdg}=0.35$ and $\lambda_\mathrm{bal}=1.1\times10^{-2}$. The relative scales are set so that $\widetilde{\mathcal{L}}_\mathrm{DSM}$ dominates the gradient throughout training. Meanwhile, $\mathcal{L}_\mathrm{pdg}$ supplies the supervisory signal to the charge head, while $\mathcal{L}_\mathrm{bal}$ acts as a weak prior whose only role is to break the residual sign ambiguity between $e^+$ and $e^-$ particles at the event level.

\subsection{Network Architecture}

The denoising network $(\boldsymbol{\eps}_\theta,\nu_\theta)$ is a permutation-equivariant transformer with shared trunk and two output heads (Fig.~\ref{fig:arch}).

\paragraph{Embeddings.} Sinusoidal time embeddings of dimension $d_\mathrm{model}=512$~\cite{Vaswani:2017},
\begin{equation}
\begin{aligned}
\phi_t(t)_{2j}&=\sin\left(\frac{t}{10^{42j/d_\mathrm{model}}}\right),\\
\phi_t(t)_{2j+1}&=\cos\left(\frac{t}{10^{42j/d_\mathrm{model}}}\right),
\end{aligned}
\end{equation}
are passed through a two-layer MLP with SiLU activation. The continuous state is linearly projected, $\phi_x(\bx_t)=W_x\bx_t$. The charge label is embedded with a learned table $\phi_k\in\mathbb{R}^{C\times d_\mathrm{model}}$, and likewise for the process label, $\phi_\pi\in\mathbb{R}^{3\times d_\mathrm{model}}$. A per-event log-multiplicity embedding, defined as:
\begin{equation}
\phi_K(K) = \mathrm{MLP}\bigl(\log K\bigr),
\end{equation}
is broadcast across the sequence. The input embedding to the transformer is then:
\begin{equation}
\mathbf{h}^{(0)}_{n,i}=\phi_x(\bx_{t,n,i})+\phi_t(t)+\phi_k(k_{t,n,i})+\phi_\pi(\pi_{n,i})+\phi_K(K_n),
\end{equation}
with the masked rows zeroed.

\paragraph{Trunk.} The trunk is a stack of $L=3$ pre-norm transformer encoder layers~\cite{Vaswani:2017,Xiong:2020} with $h=2$ heads, hidden dimension $d_\mathrm{model}=512$, feed-forward expansion $4d_\mathrm{model}$, GELU activation and dropout $p=0.085$. Padding slots are excluded by a key-padding mask. A residual skip with coefficient $\alpha_\mathrm{skip}=0.2$ between the input embedding and the trunk output stabilizes training at large $L$.

\paragraph{Heads.} Two linear heads operate on the trunk output:
\begin{align}
\boldsymbol{\eps}_\theta &= W_\mathrm{out}\mathbf{h}^{(L)}\quad\in\mathbb{R}^7,\\
\nu_\theta &= W_\mathrm{pdg}\mathbf{h}^{(L)}\quad\in\mathbb{R}^2.
\end{align}
The continuous output is masked. The total trainable parameter count is approximately $1.7\times10^7$.

\subsection{Training}
\label{sec:training}

The training loop minimizes Eq.~\eqref{eq:Ltot} over the standardized dataset using the \textsc{AdamW}~\cite{Loshchilov:2017adamw} optimizer with a weight decay of $10^{-4}$. We employ the OneCycle learning-rate schedule~\cite{Smith:2017onecycle}, with peak learning rate $\eta_{\max}=1.6\times10^{-4}$, warm-up fraction $p_s=0.072$, initial division factor $D=30.9$ and final division factor $F=10^4$. Gradients are clipped to $\|\nabla\|_2\le 4.59$. The model is trained for $20$ epochs at batch size $B=8$ (events) on a single NVIDIA A100 GPU with TF32 enabled and cuDNN deterministic kernels selected. Algorithm~\ref{alg:train} summarizes one training step. The canonical \texttt{PyTorch} deep learning library is used for all implementation. 

\begin{algorithm}[H]
\caption{One training step of {\deffusion}.}
\label{alg:train}
\begin{algorithmic}[1]
\State Sample event batch $\{(\bx_0,k_0,\pi_0,m)\}_{n=1}^B$ from $\mathcal{D}_\mathrm{train}$.
\State Sample $t\sim\mathrm{Unif}\{1,\ldots,T\}$, $\boldsymbol{\eps}\sim\mathcal{N}(0,\mathbf{I})$.
\State $\bx_t\gets \sqrt{\overline\alpha_t}\bx_0+\sqrt{1-\overline\alpha_t}\boldsymbol{\eps}$ %\Comment{Eq.~\eqref{eq:reparam}}
\State $k_t\sim q(\cdot|k_0,t)$ %\Comment{uniform-flip kernel}
\State $\pi_t\gets \pi_0$ %\Comment{clean conditioning}
\State $\boldsymbol{\eps}_\theta,\nu_\theta\gets \mathcal{M}_\theta(\bx_t,t,k_t,\pi_0,m)$
\State Compute $\widetilde{\mathcal{L}}_\mathrm{DSM}, \mathcal{L}_\mathrm{pdg}, \mathcal{L}_\mathrm{bal}$.
\State $\theta\gets \theta-\eta(\tau)\nabla_\theta\mathcal{L}$.
\end{algorithmic}
\end{algorithm}

\subsection{Sampling}
\label{sec:sampling}

Given a trained model, drawing a synthetic bunch crossing proceeds in three stages.

\paragraph{Multiplicity and composition.} Sample a real index $n^\star\sim\mathrm{Unif}\{1,\ldots,N_\mathrm{ev}\}$. Set the target multiplicity $K=K_{n^\star}$ and the per-event composition $\boldsymbol{f}=\boldsymbol{f}_{n^\star}$. Build the mask $m_i=\mathbbm{1}[i<K]$.

\paragraph{Initial state.} Sample the prior:
\begin{align}
\bx_T &\sim\mathcal{N}(0,\mathbf{I}) \odot m,\\ % was: \odotm,\\
k_T   &\sim\mathrm{Unif}\{0,1\} \odot m,\\ % was: \odotm,\\
\pi_i &\sim\mathrm{Cat}(\boldsymbol{f})\quad\text{for } i<K.
\end{align}
The process labels are drawn once and held fixed for all $t$.

\paragraph{Reverse iteration.} For $t=T,T-1,\ldots,1$:
\begin{enumerate}
\item Compute $\boldsymbol{\eps}_\theta,\nu_\theta=\mathcal{M}_\theta(\bx_t,t,k_t,\pi,m)$.
\item Apply Eq.~\eqref{eq:reverse_cont}, then clip to $\|\bx_{t-1}\|_\infty\le 10$.
\item Sample $k_{t-1}\sim\mathrm{Cat}(\sigma(\nu_\theta))$ at masked slots.
\end{enumerate}
The decoded event is obtained by inverting the standardization, applying Eq.~\eqref{eq:squash} and Eq.~\eqref{eq:logE}, and writing:
\begin{equation}
\mathcal{E}_\mathrm{gen} = \bigl\{(11(1-2k_i),E_i,\bbeta_i,\mathbf{r}_i,\pi_i)\bigr\}_{i<K}.
\end{equation}
Where lower fidelity samples are acceptable, we also allow the user to optionally subsample the reverse trajectory to $T'<T$ steps with linearly-spaced time steps for faster generation.

\section{Results}
\label{sec:results}

\subsection{Particle-Level Distributions}

\begin{figure*}[t]
\centering
\includegraphics[width=0.32\linewidth]{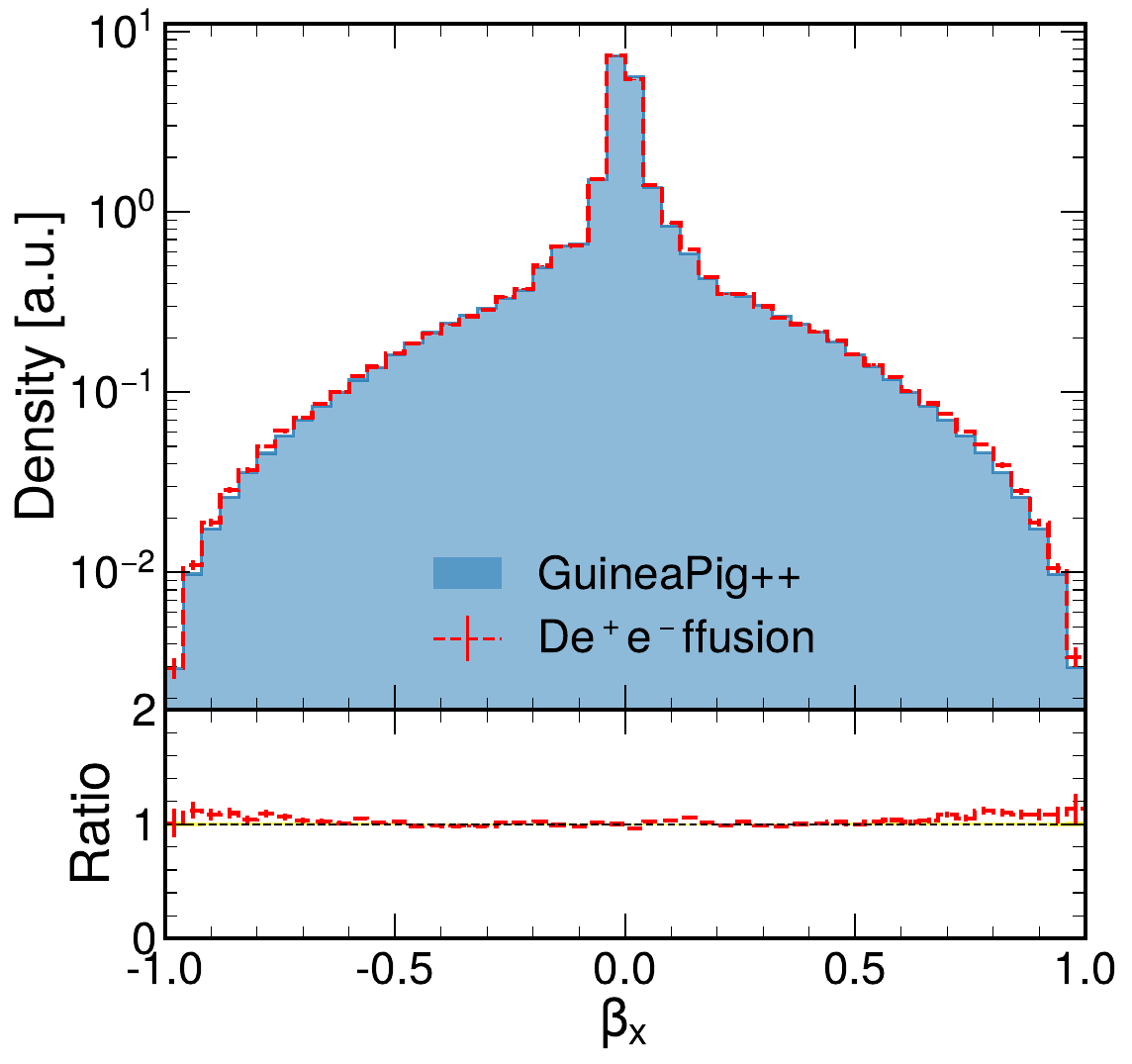}
\includegraphics[width=0.32\linewidth]{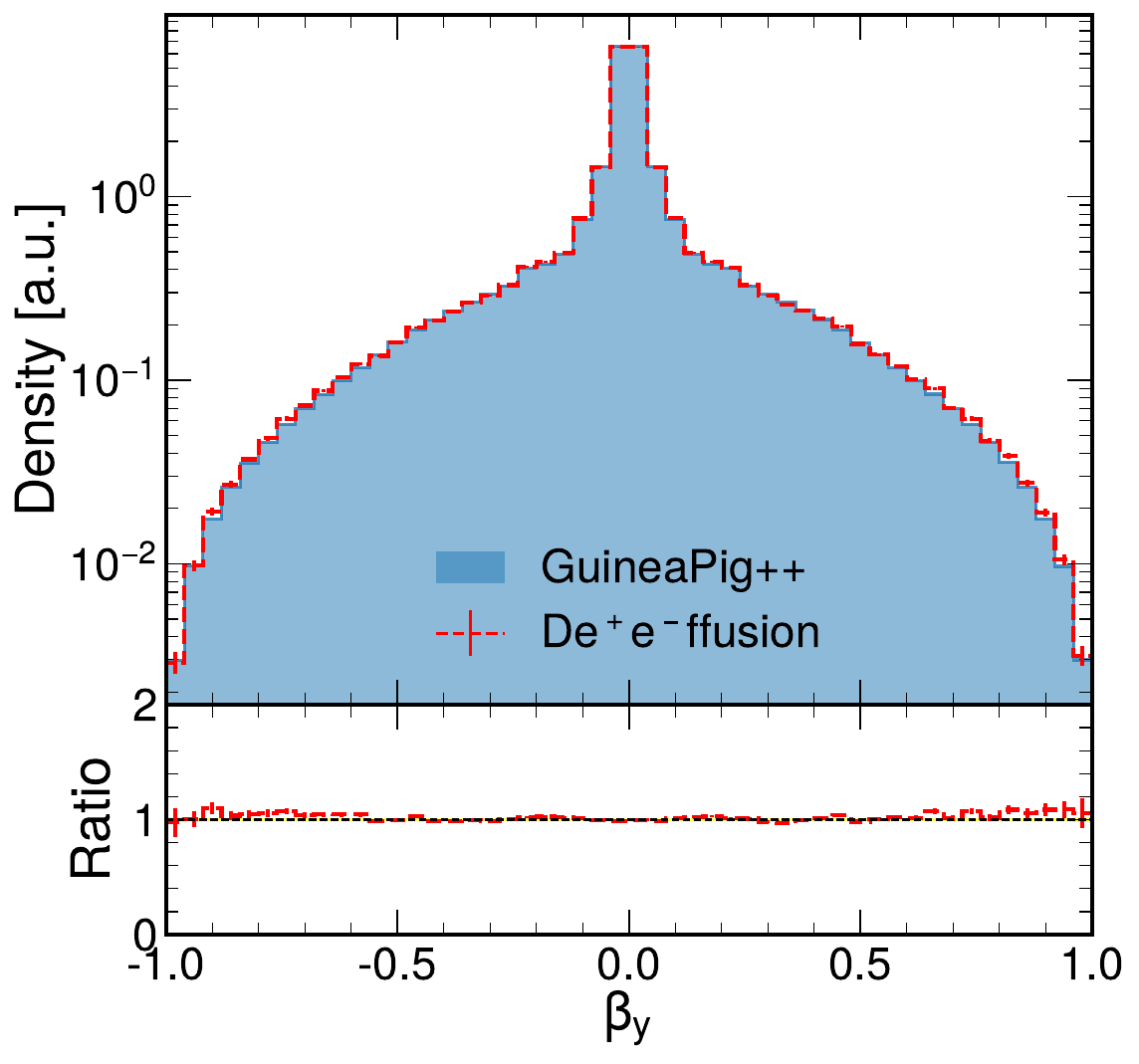}
\includegraphics[width=0.32\linewidth]{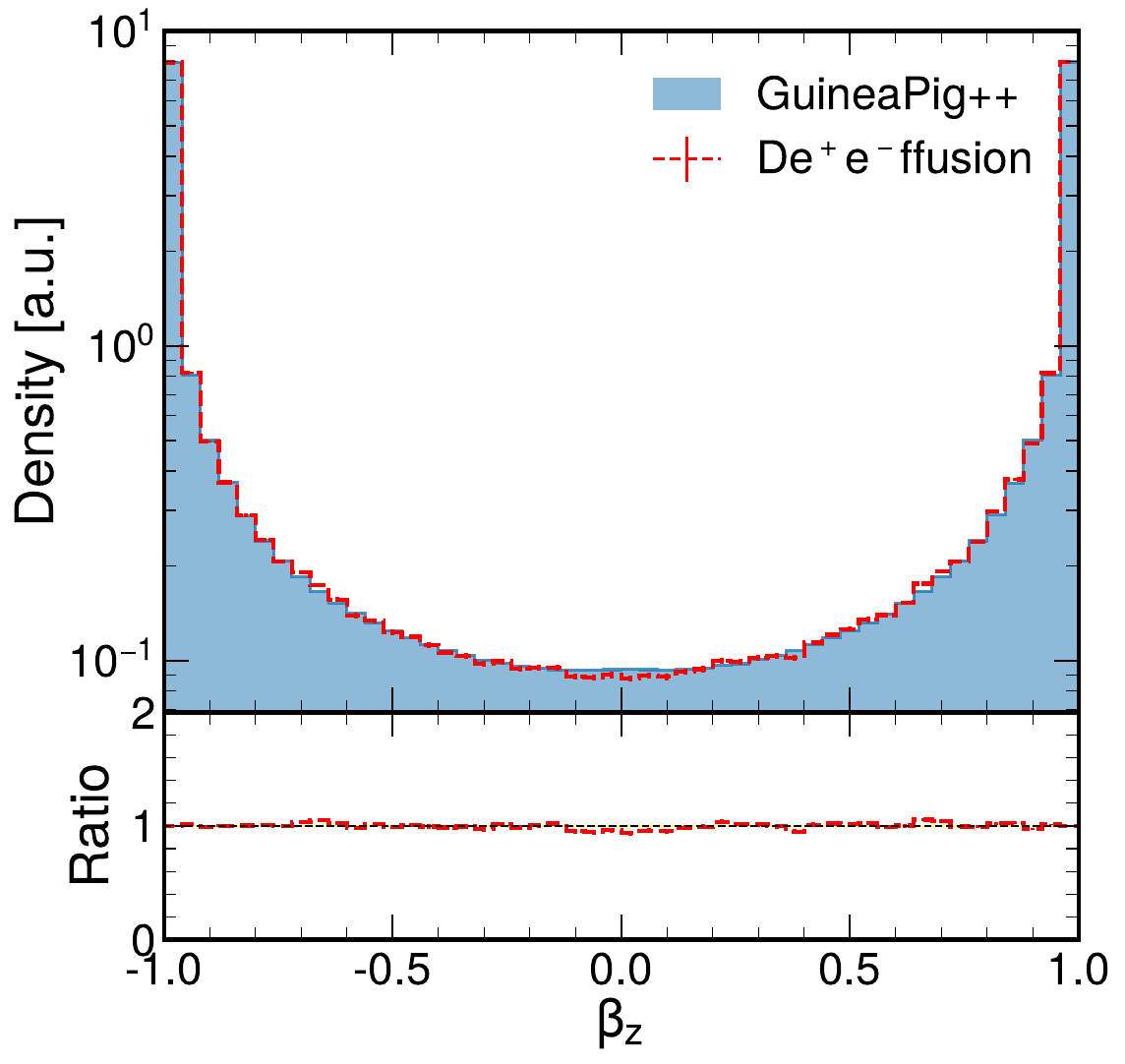}
\includegraphics[width=0.32\linewidth]{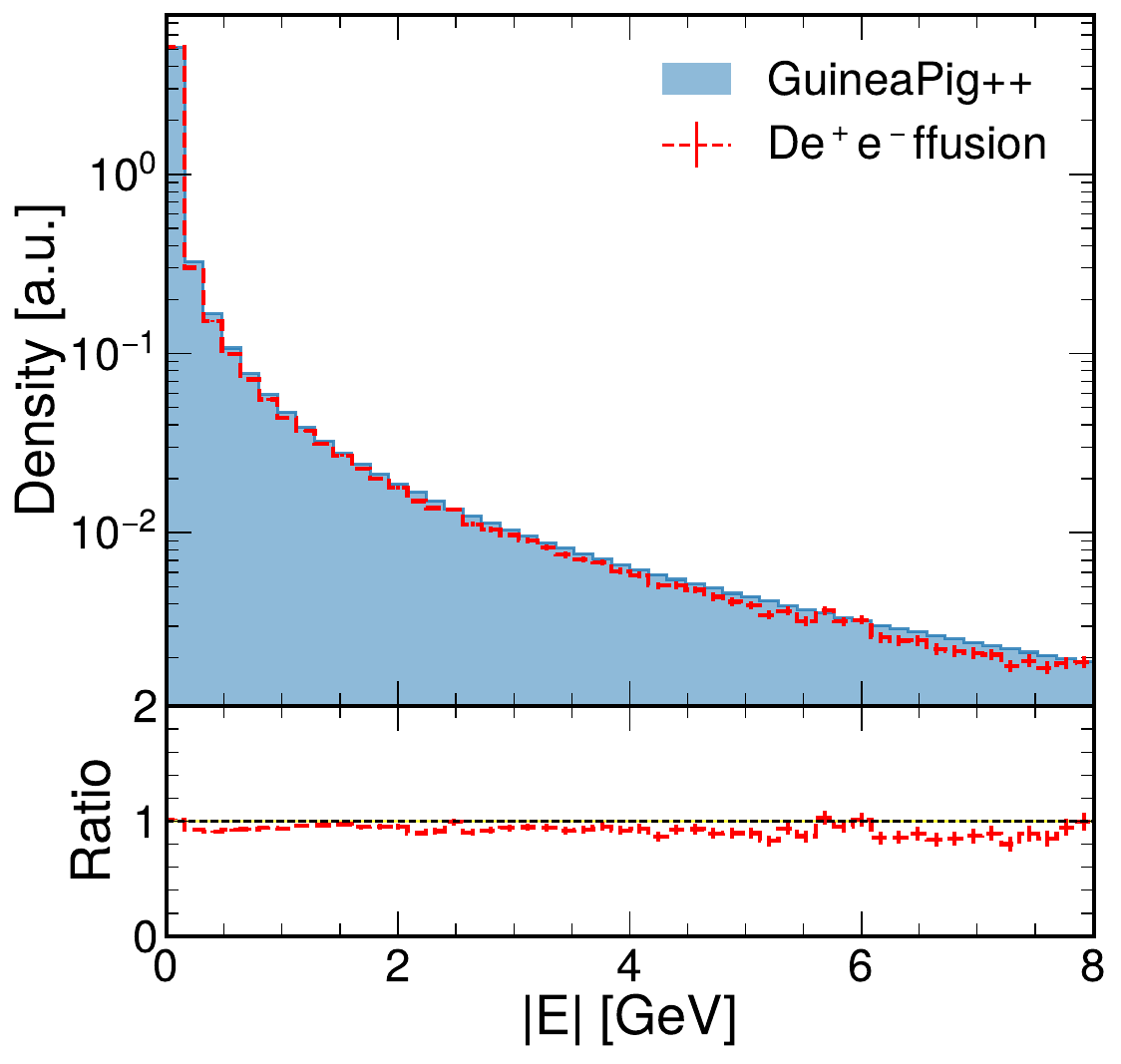}
\includegraphics[width=0.32\linewidth]{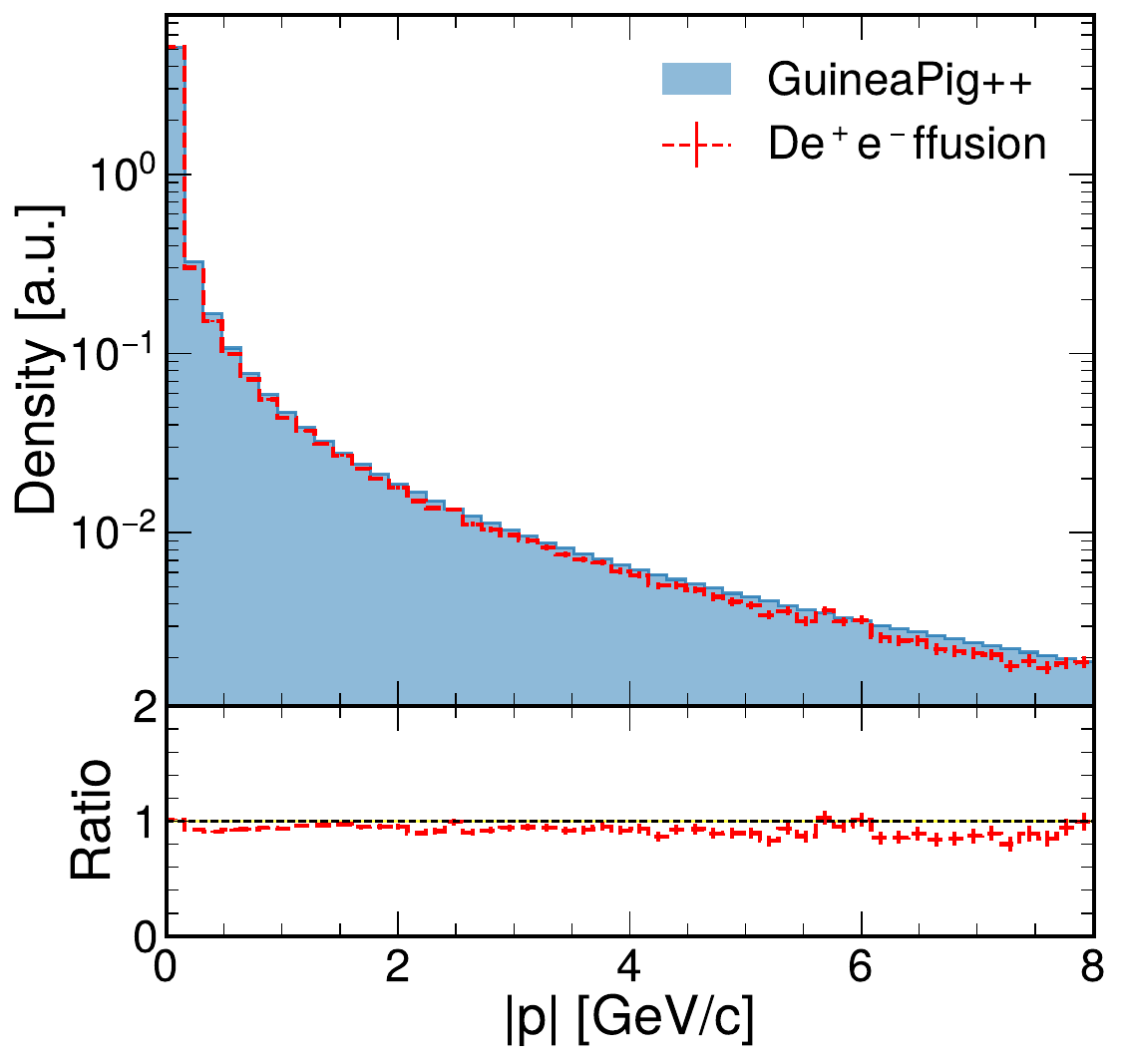}
\includegraphics[width=0.32\linewidth]{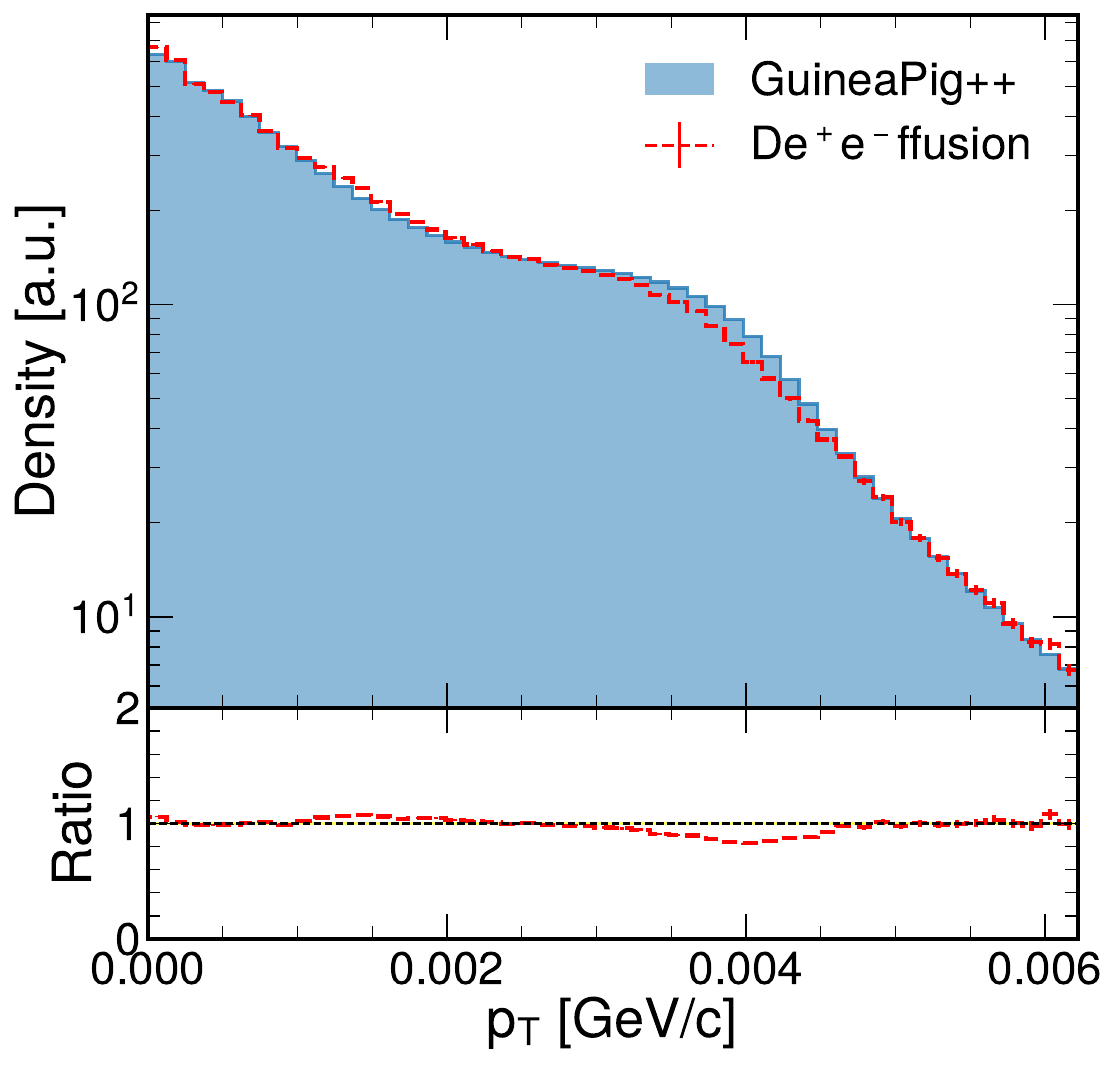}
\includegraphics[width=0.32\linewidth]{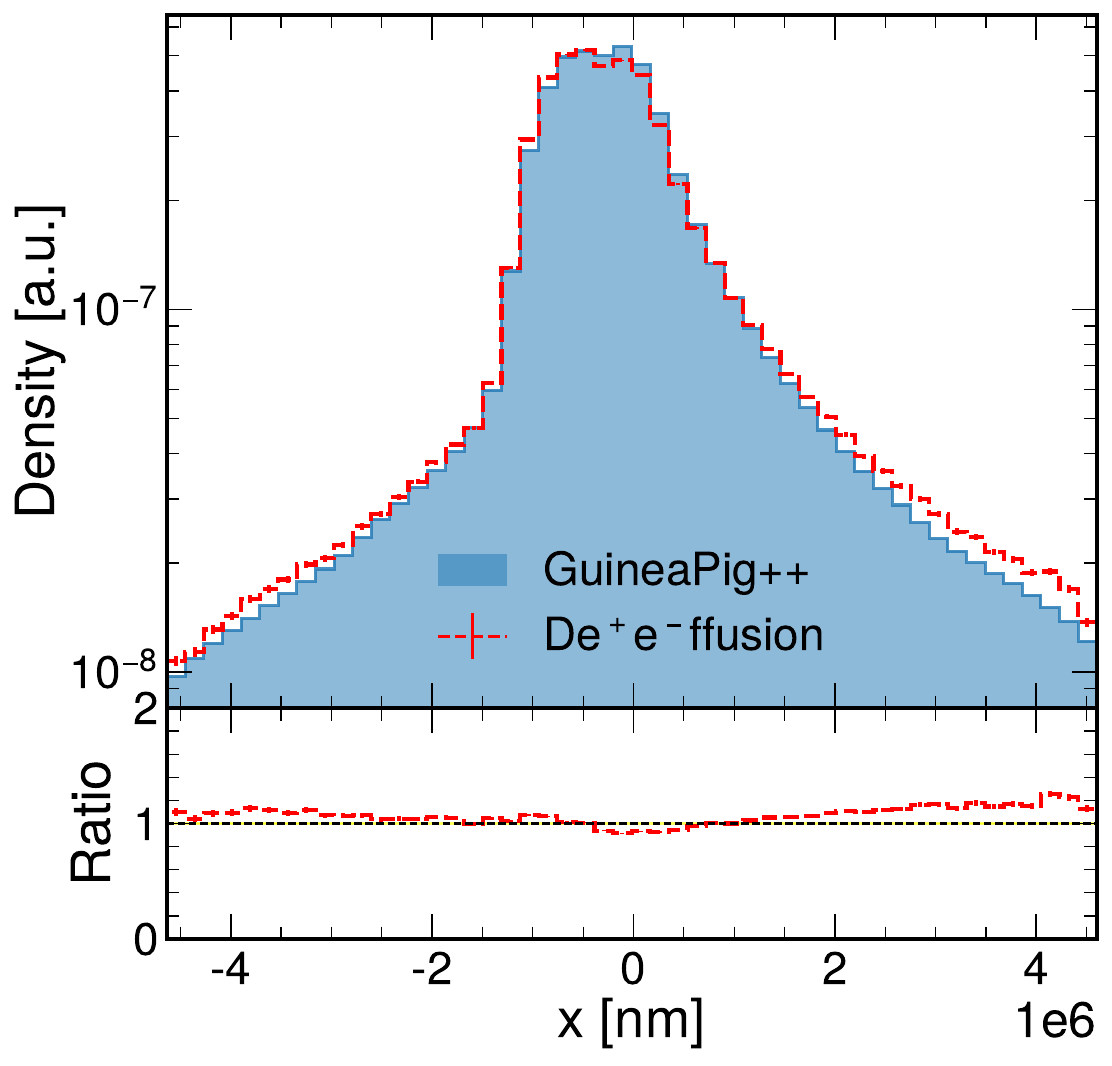}
\includegraphics[width=0.32\linewidth]{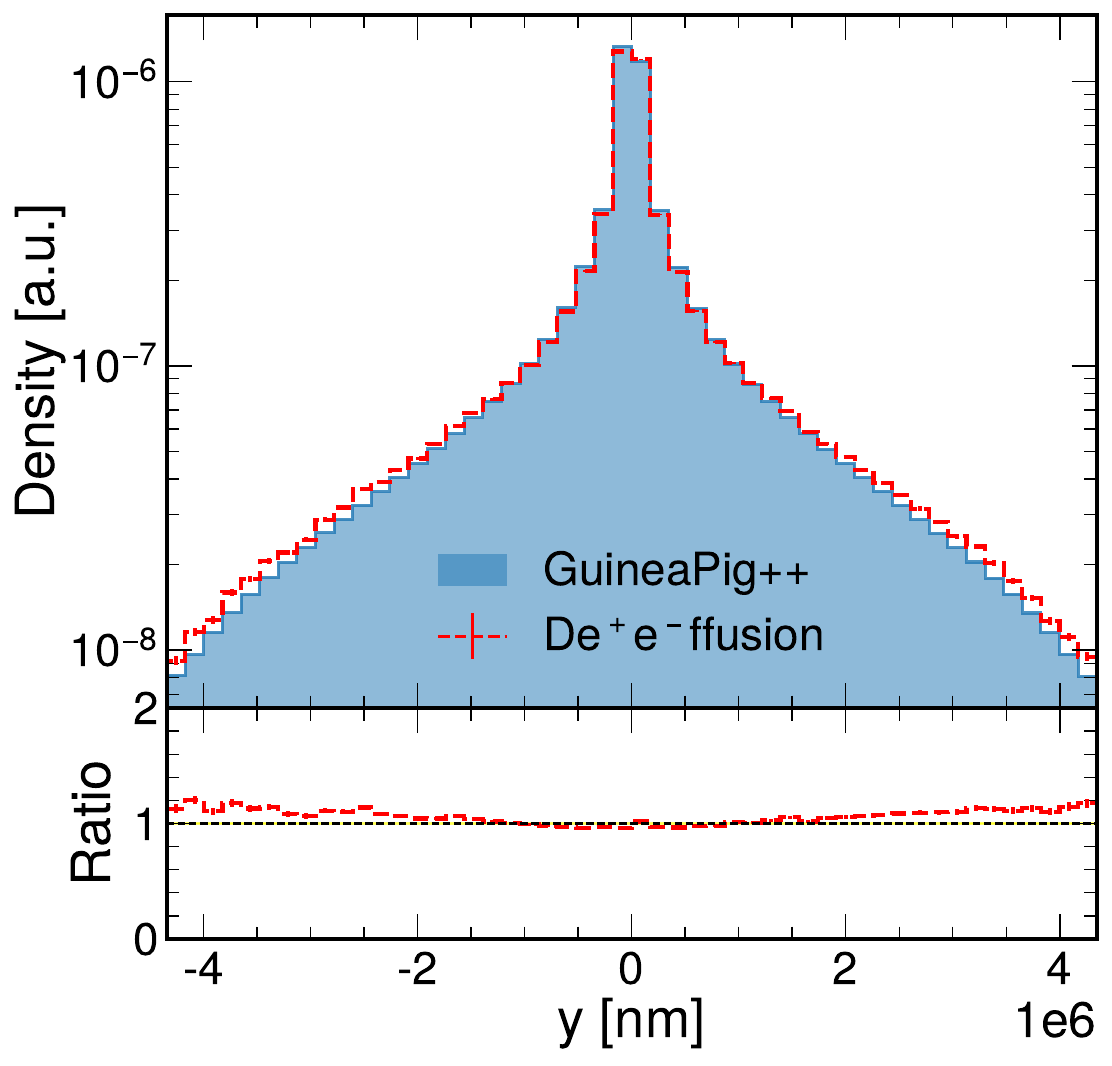}
\includegraphics[width=0.32\linewidth]{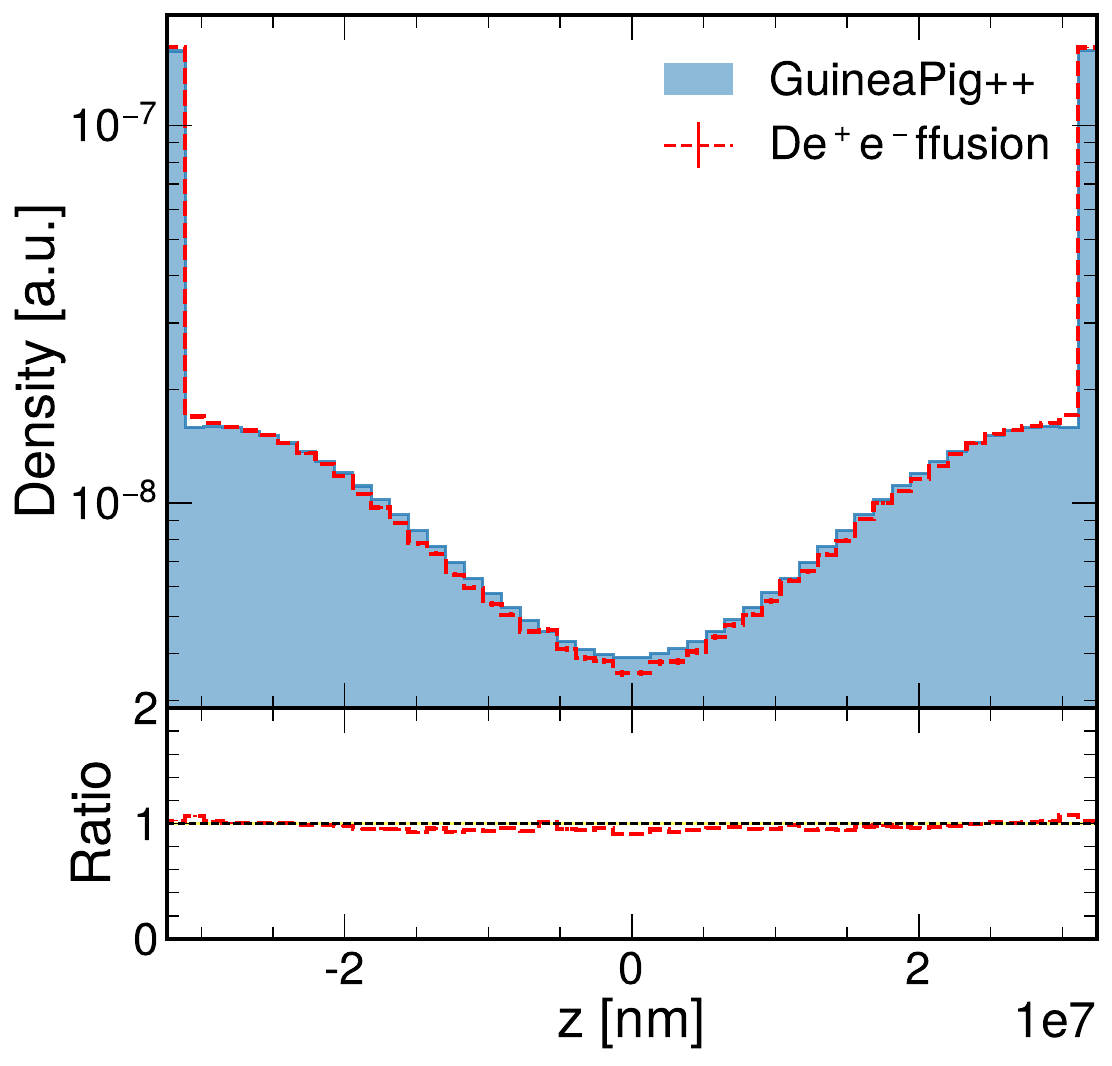}
\caption{\justifying Particle-level one-dimensional marginal distributions of $\beta_x$ (top-left), $\beta_y$ (top-middle), $\beta_z$ (top-right), $\abs{E}$ (center-left), $\abs{p}$ (center-middle), $\pT$ (center-right), $x$ (bottom-left), $y$ (bottom-middle), and $z$ (bottom-right). For each panel the filled blue histogram corresponds to {\gpig}, red dashed is the diffusion model {\deffusion}. All histograms are normalized to unit area. The error bars represent the statistical uncertainty in each bin. The bottom panels in each plot show the ratio of the {\deffusion} model's prediction of each observable to their truth {\gpig} values and the yellow-shaded region indicates the statistical uncertainty in the truth.}
\label{fig:marg}
\end{figure*}

Since the output space of events generated by both {\gpig} and {\deffusion} is too high-dimensional to be examined holistically, we first compare them at the level of one-dimensional marginals of the per-particle features. Figure~\ref{fig:marg} shows the comparison of $|E|$, $|\mathbf{p}|$, $\pT$, the three components of $\bbeta$, and the three components of the position, separately for $e^+$ and $e^-$. The two models agree at the percent level within their statistical uncertainties across the bulk of every distribution; the heavy tails of $|E|$ and $|\pT|$ are also recovered.

The velocity components, $\beta_x$ and $\beta_y$, shown in the top row of Fig.~\ref{fig:marg}, are sharply peaked near zero with approximately symmetric tails extending to $|\beta_{x,y}|\simeq 1$. {\deffusion} reproduces both the narrow central enhancement and the broad falloff into the tails, with the ratio panels remaining close to unity over essentially the full range. The longitudinal component $\beta_z$ has a qualitatively different structure, with a strong enhancement near $|\beta_z|\simeq 1$ and a depleted central region around $\beta_z\simeq 0$, reflecting the highly forward/backward nature of the beamstrahlung pair background. This structure is also well modeled, within statistical uncertainty, including the steep rise near the kinematic endpoints.

The middle row of Fig.~\ref{fig:marg} displays the energy and momentum spectra, $|E|$ and $|p|$, which fall steeply over several orders of magnitude from low values up to ${\sim}8$\,GeV. {\deffusion} follows the {\gpig} prediction throughout the bulk of these distributions, accurately capturing both the large low-energy population and the rapidly decreasing high-energy tail. The transverse momentum distribution, $\pT$, is especially steep and concentrated at very small values, with most of the density below a few MeV. The generated spectrum tracks the reference distribution closely, with deviations at the 10\% level at the inflection point around 4\,MeV, where there is a change caused by a moderate bimodal structure. Further small deviations are mainly confined to the lowest density bins, where the distributions are most sensitive to statistical fluctuations and binning effects.

Finally, the bottom row shows the spatial distributions in $x$ and $y$, which exhibit compact transverse beam profiles with central peaks and long non-Gaussian tails extending over several $10^6$\,nm. {\deffusion} reproduces the peak positions, widths, and tail behavior in both transverse directions, indicating that the model has learned not only the momentum-space structure but also the spatial beam profile. The longitudinal coordinate $z$ spans a much larger scale, of order $10^7$\,nm, and exhibits a broad, edge-enhanced distribution with a minimum near the center. The generated distribution again remains in close agreement with {\gpig}, with only mild residual differences visible in the tails. Overall, the near-flat ratio panels demonstrate that {\deffusion} accurately reproduces the full one-dimensional phase-space marginals of {\gpig}, including sharply peaked, highly asymmetric, and multi-scale features across velocity, energy-momentum, and spatial coordinates.

\subsection{Per-Process Longitudinal Distribution}

\begin{figure*}[t]
\centering
\includegraphics[width=0.32\linewidth]{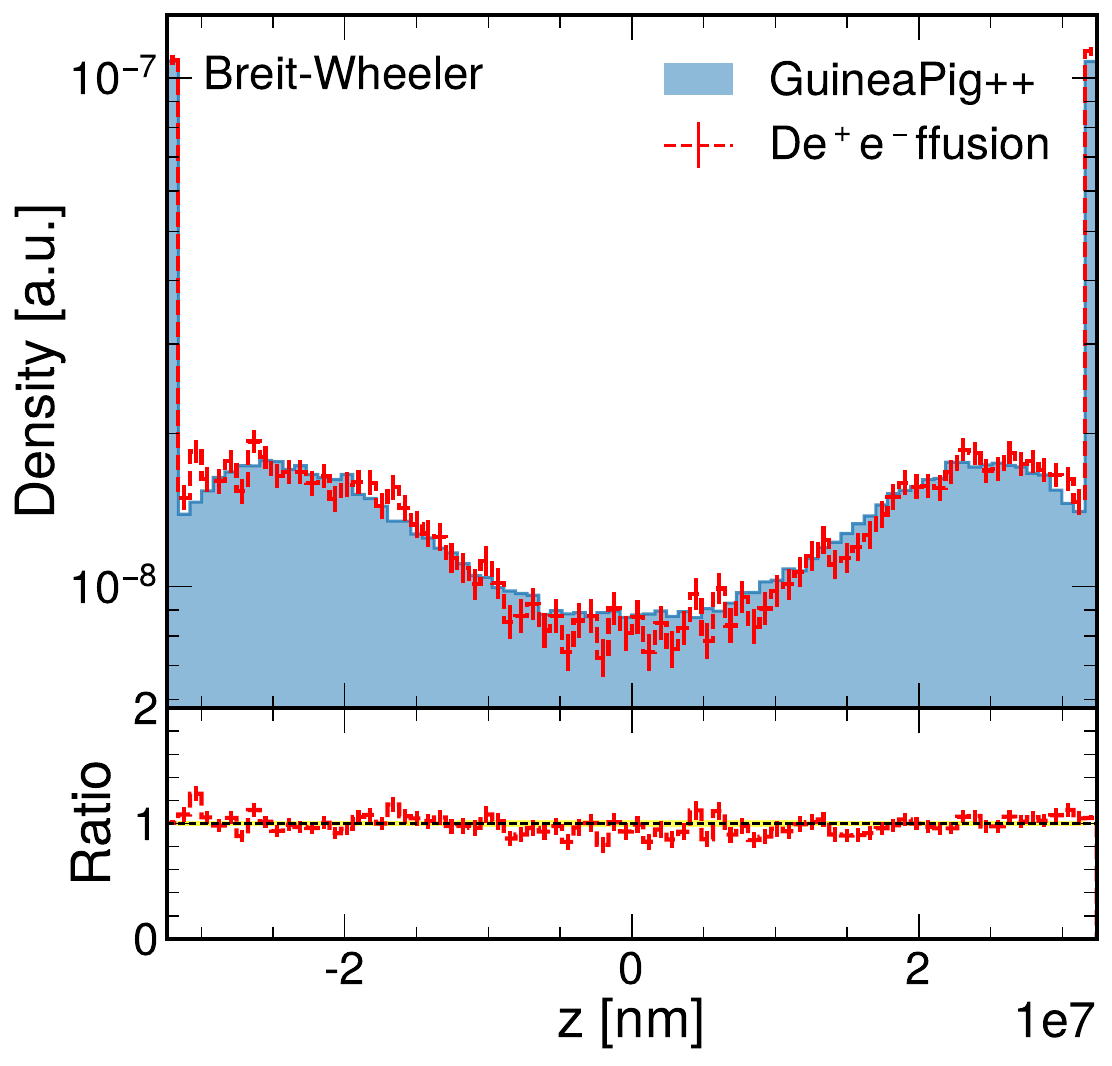}
\includegraphics[width=0.32\linewidth]{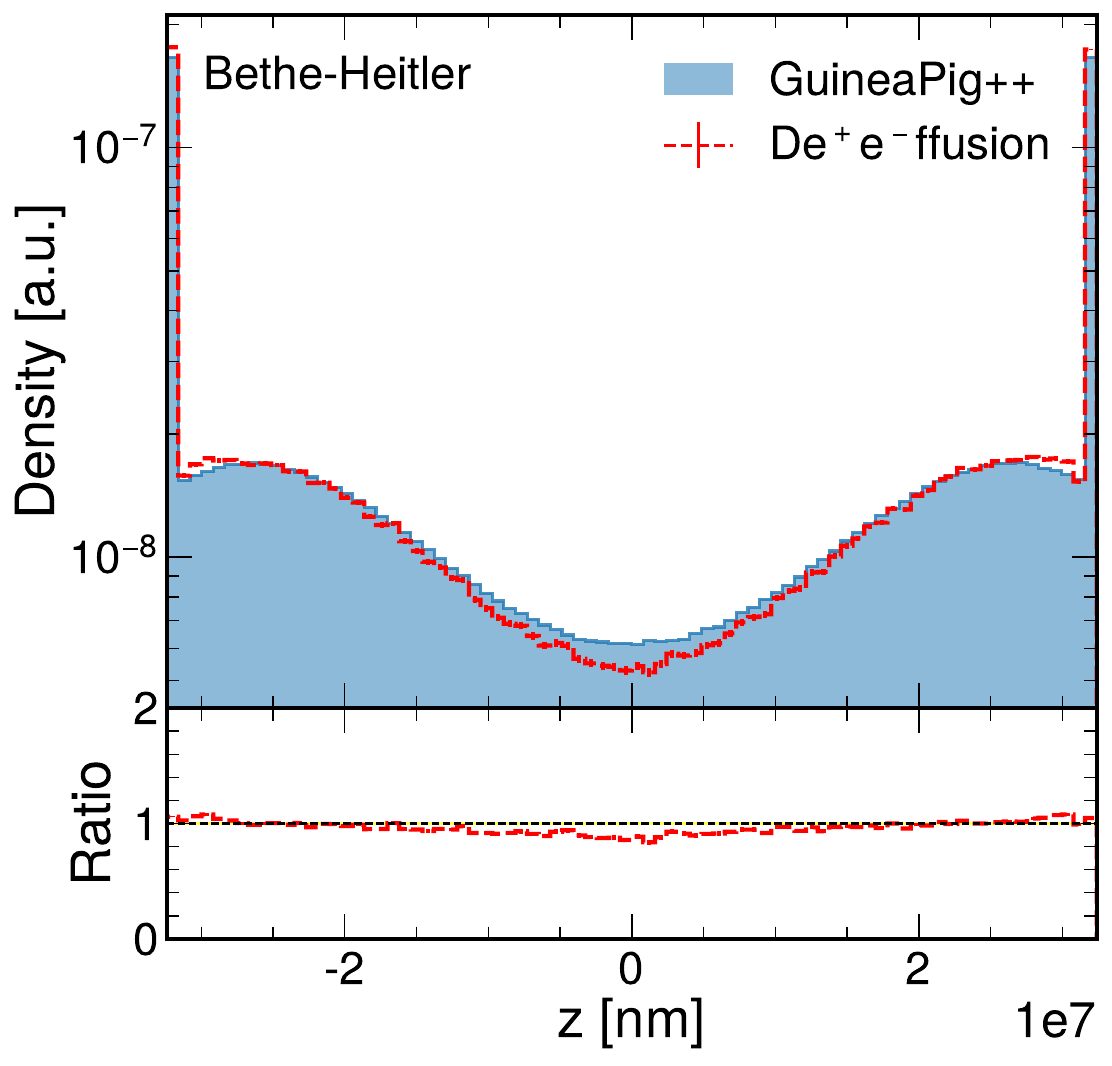}
\includegraphics[width=0.32\linewidth]{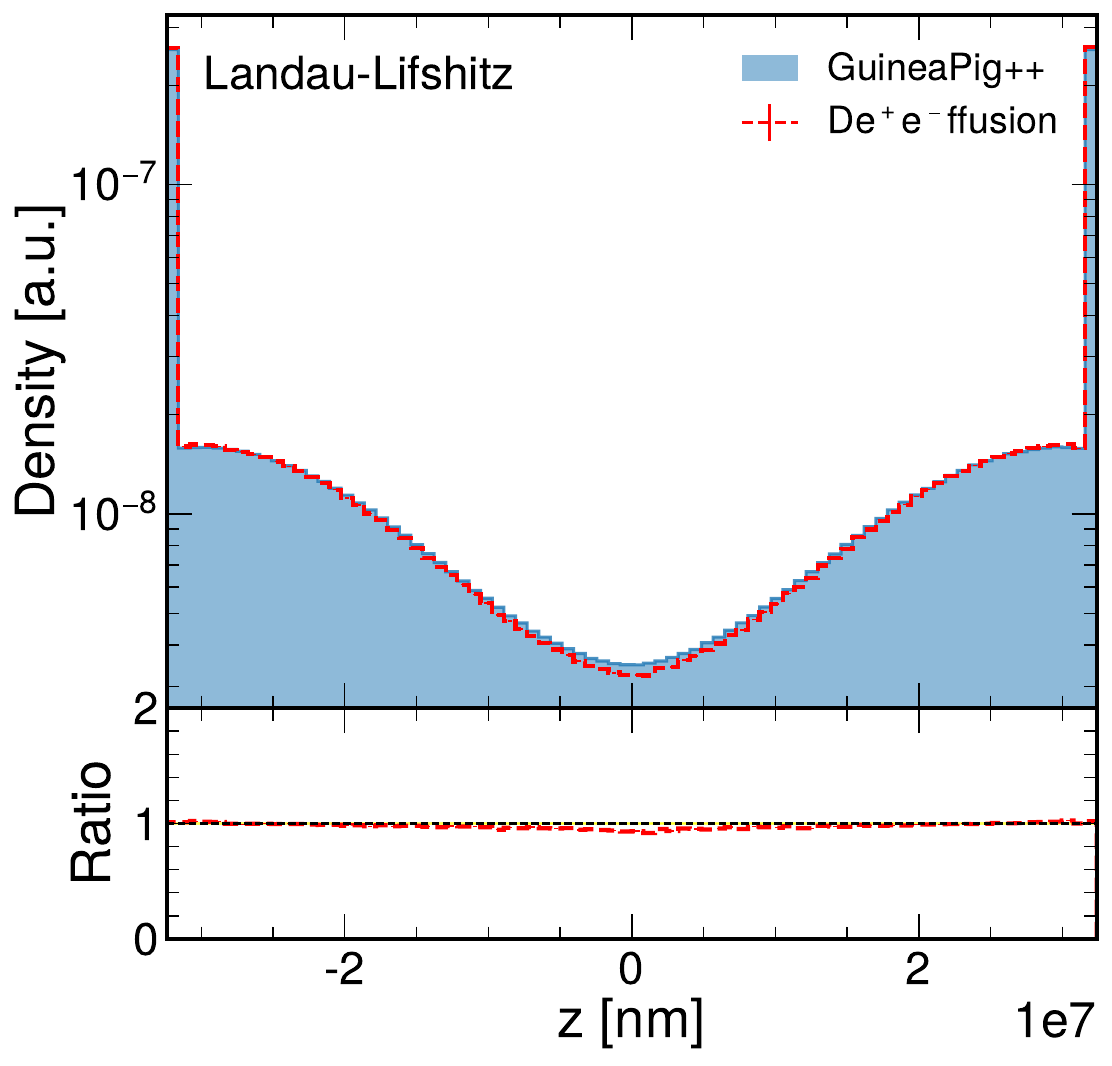}
\caption{\justifying Longitudinal $z$ position distribution, split by QED subprocess Breit--Wheeler (left), Bethe--Heitler (center), Landau--Lifshitz (right) for {\gpig} (filled blue) and {\deffusion} (red dashed). All histograms are normalized to unit area. The error bars represent the statistical uncertainty in each bin. The ratio of the {\deffusion} model's prediction to the truth {\gpig} value is shown in the bottom panel. The yellow-shaded region indicates the statistical uncertainty in the truth.}
\label{fig:zproc}
\end{figure*}

Another stringent particle-level test is the longitudinal position coordinate $z$ split by QED subprocess (Fig.~\ref{fig:zproc}). The unconditional model averages the three channels and produces a single broad central peak; the nominal model presented in this study with process conditioning resolves the particle $z$ distribution differentially by process.

The three panels compare the longitudinal $z$ distributions for Breit--Wheeler, Bethe--Heitler, and Landau--Lifshitz pair production as predicted by {\gpig} and {\deffusion}. In all three cases, the distributions are broad over scales of order $10^7$\,nm and exhibit enhanced density in the forward and backward regions, with a depleted region around $z \simeq 0$. {\deffusion} reproduces this longitudinal structure well, with the ratio panels remaining close to unity across nearly the entire plotted range.

For Breit--Wheeler production, the $z$ spectrum is comparatively flatter and also exhibits more statistical noise, due to the significantly lower BW cross section, with modest structure in the central region and stronger statistical fluctuations in the tails. The generated distribution follows the {\gpig} reference throughout the bulk of the spectrum, including the shallow minimum near the center and the rise toward large $|z|$. The deviations are mostly localized to individual bins and are consistent with the lower effective statistics in this channel.

The Bethe--Heitler and Landau--Lifshitz panels show a smoother and more clearly U-shaped longitudinal profile, with densities lowest near $z=0$ and increasing toward both positive and negative $z$. {\deffusion} captures both the central depletion and the edge enhancement, indicating that it learns the process-dependent bunch-position structure rather than only reproducing an average profile. The agreement is especially strong for Landau--Lifshitz, where the ratio remains nearly flat at unity across the full range. Overall, these comparisons show that {\deffusion} accurately models the longitudinal production profile for all three pair-production mechanisms, with only small residual mismodeling in statistically sparse edge or low-density bins.

\subsection{Particles in Fiducial Volume}

\begin{figure*}[t]
\centering
\includegraphics[width=0.32\linewidth]{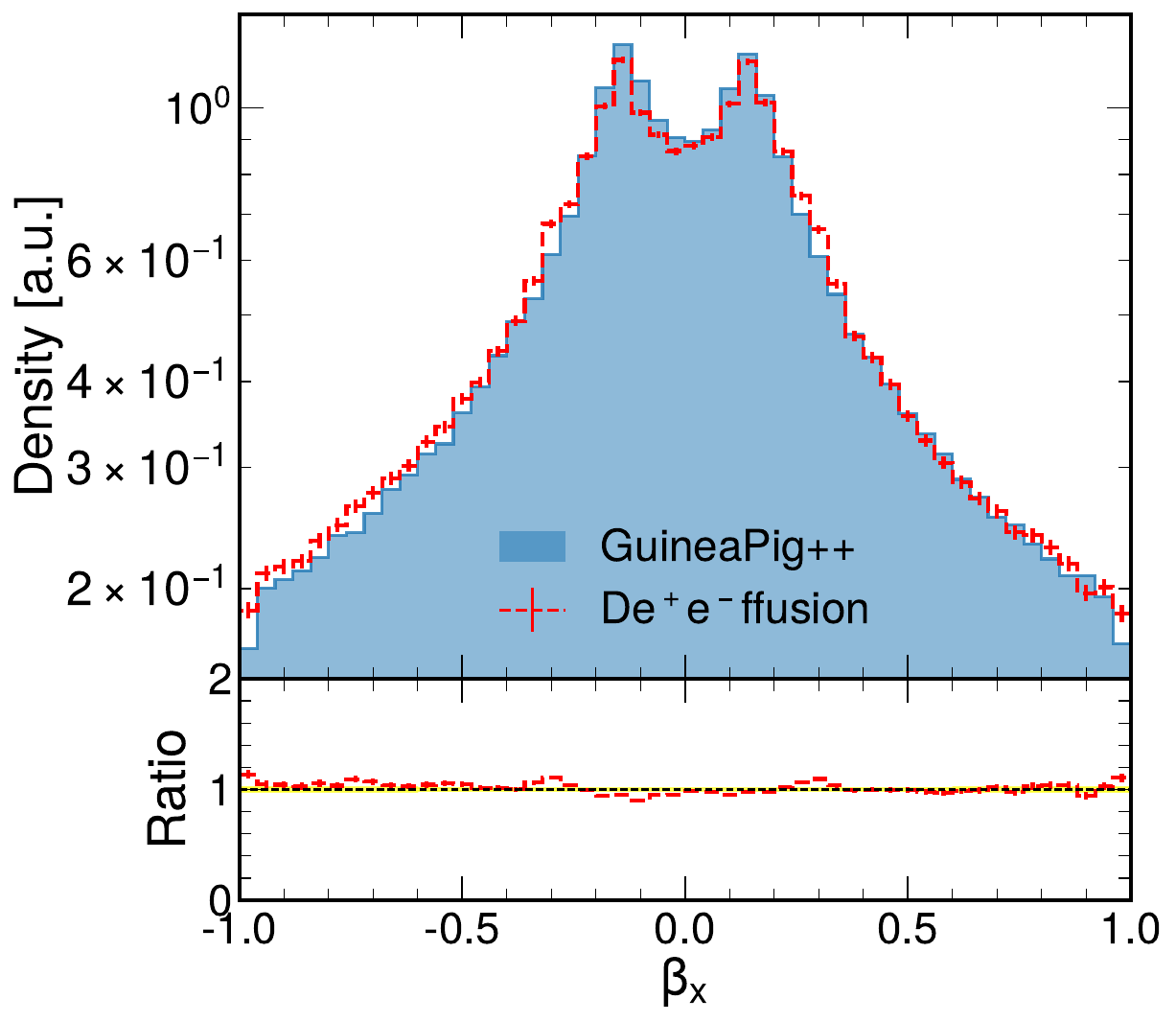}
\includegraphics[width=0.32\linewidth]{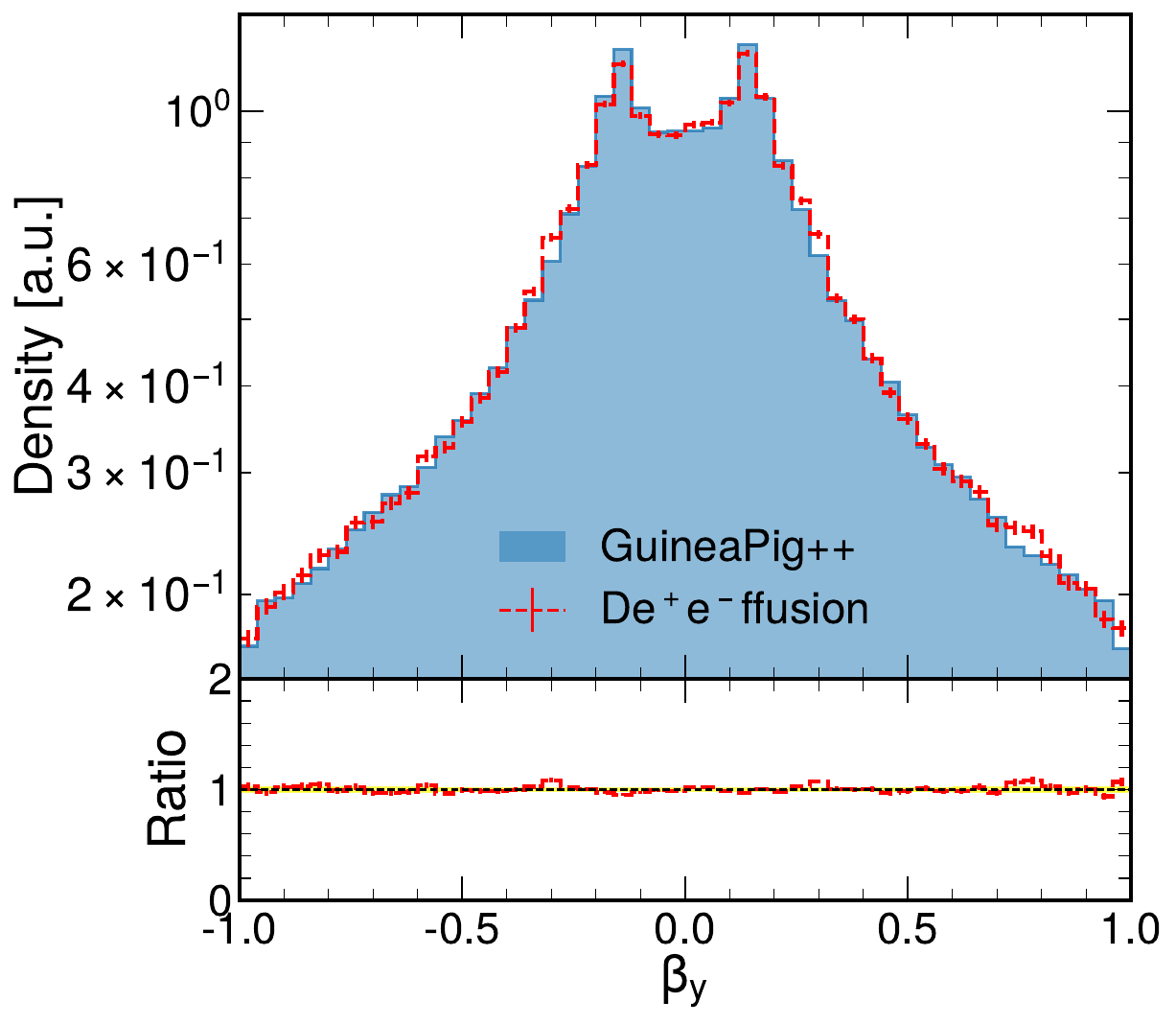}
\includegraphics[width=0.32\linewidth]{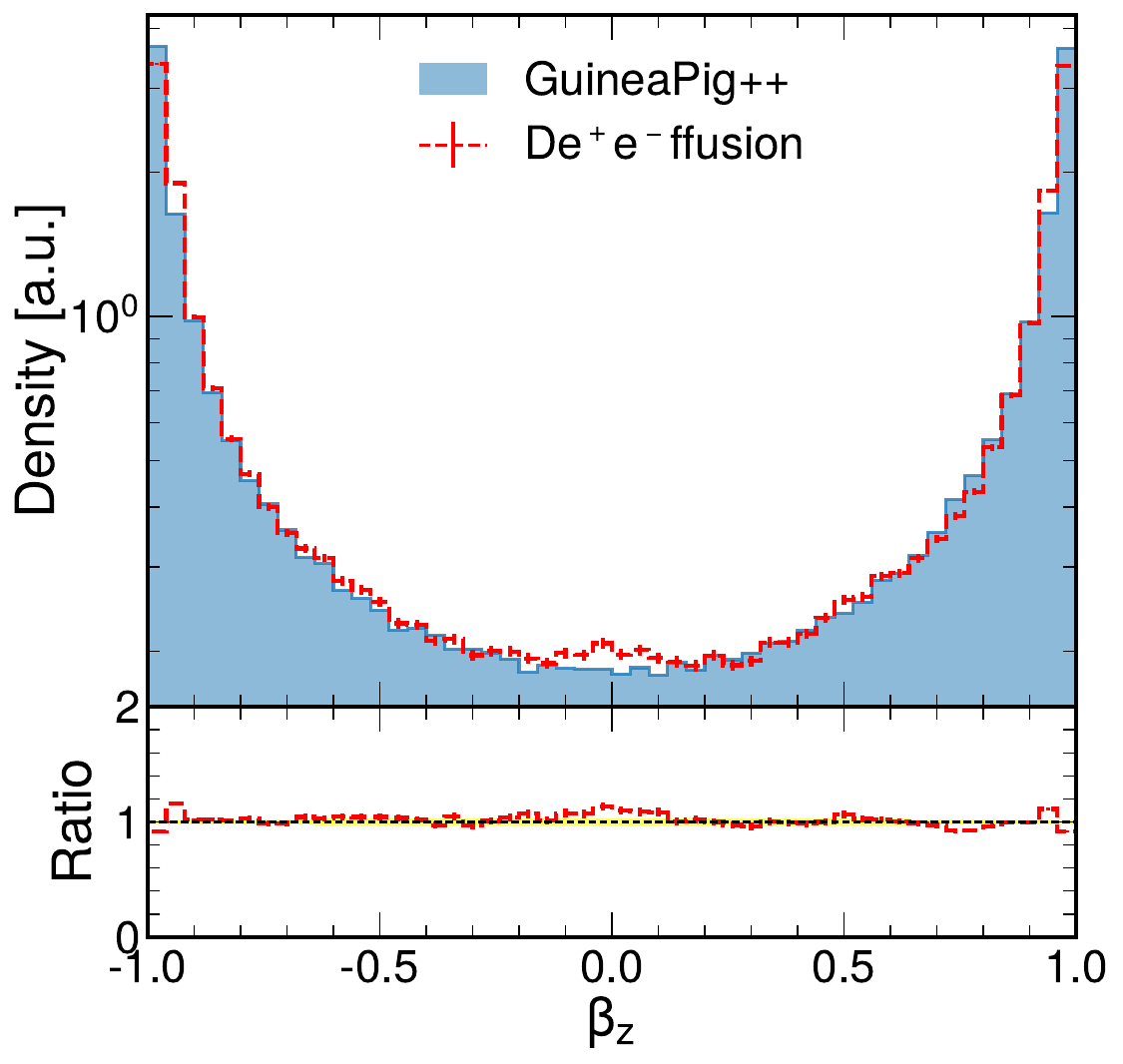}
\includegraphics[width=0.32\linewidth]{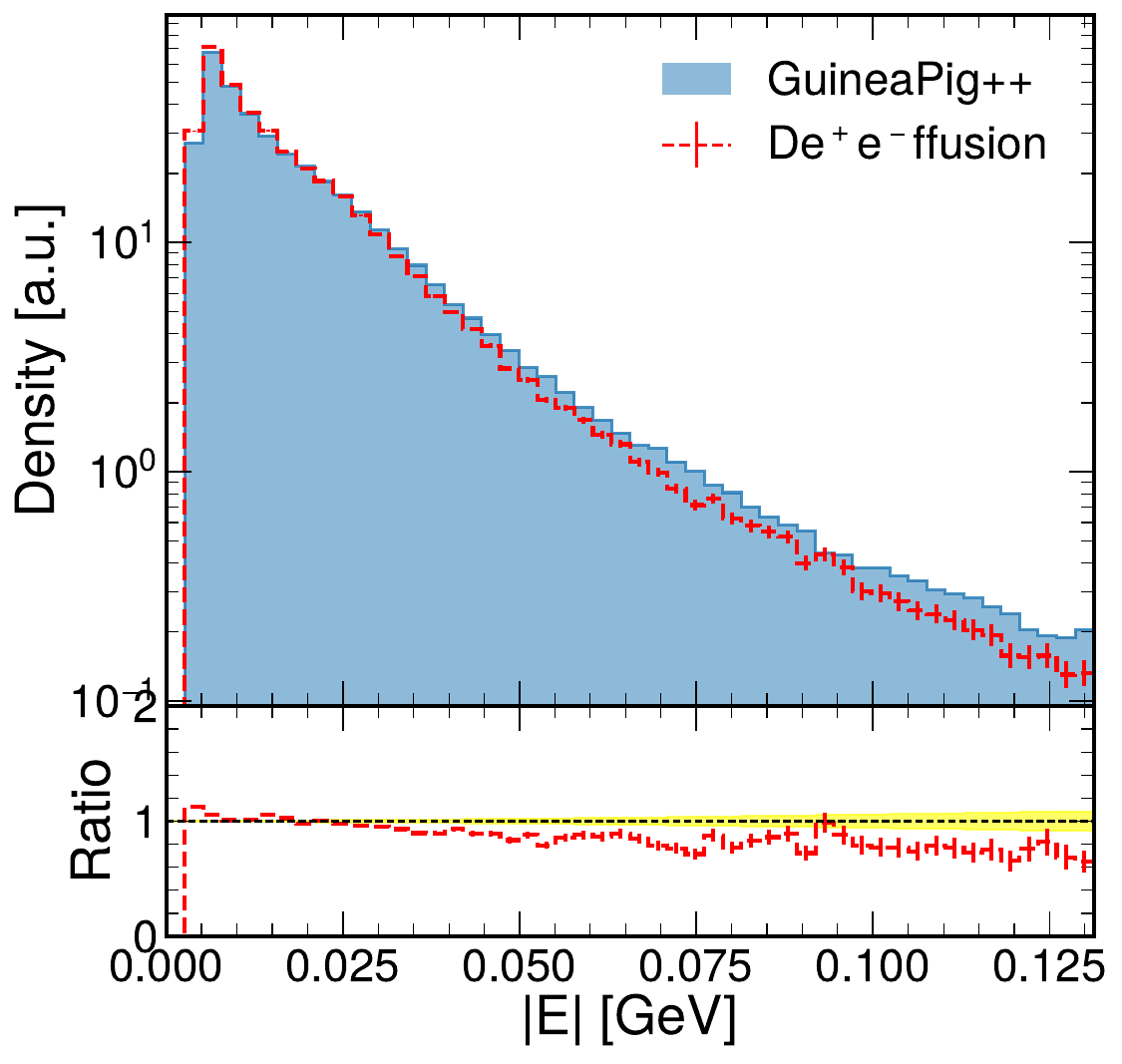}
\includegraphics[width=0.32\linewidth]{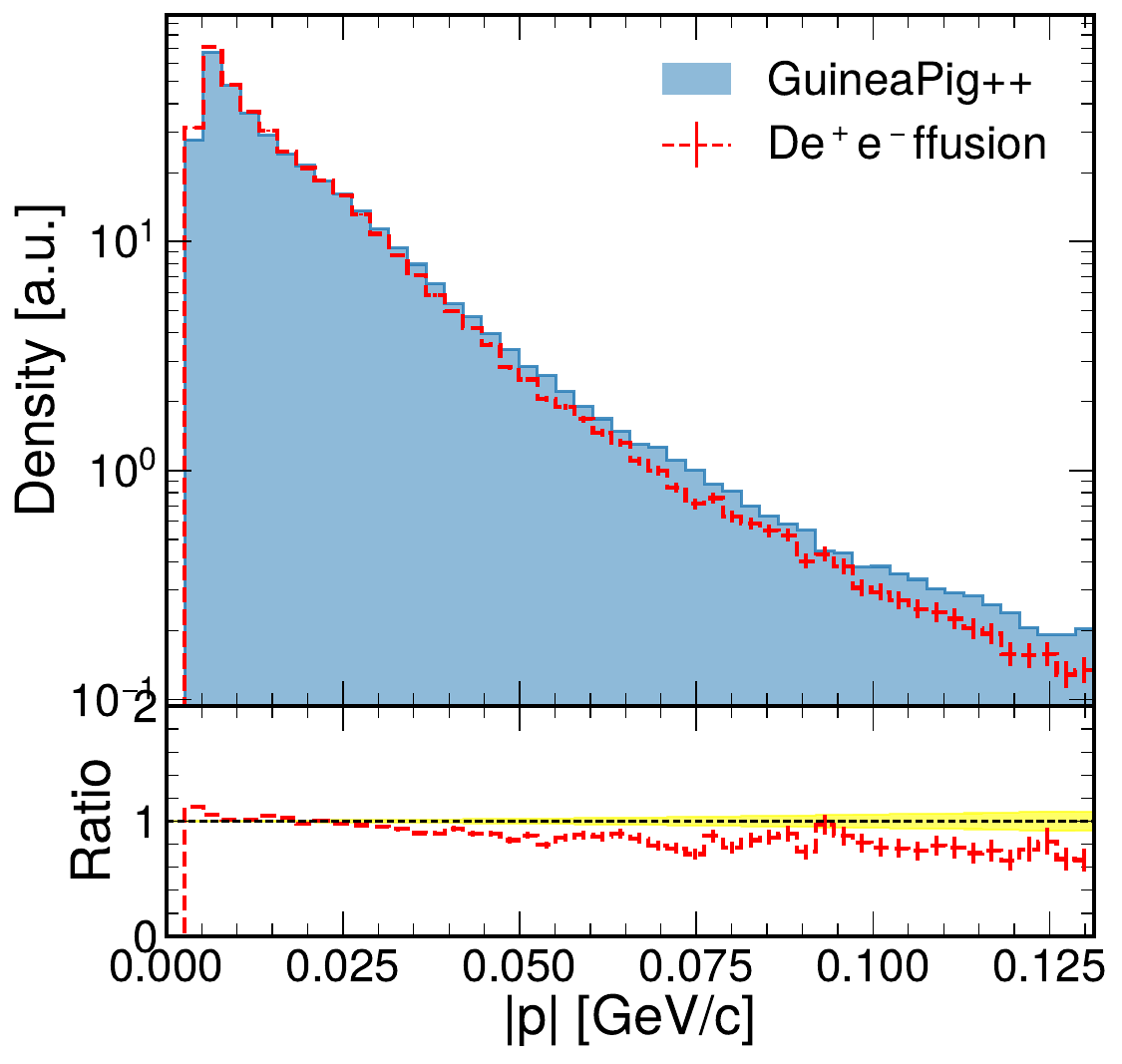}
\includegraphics[width=0.32\linewidth]{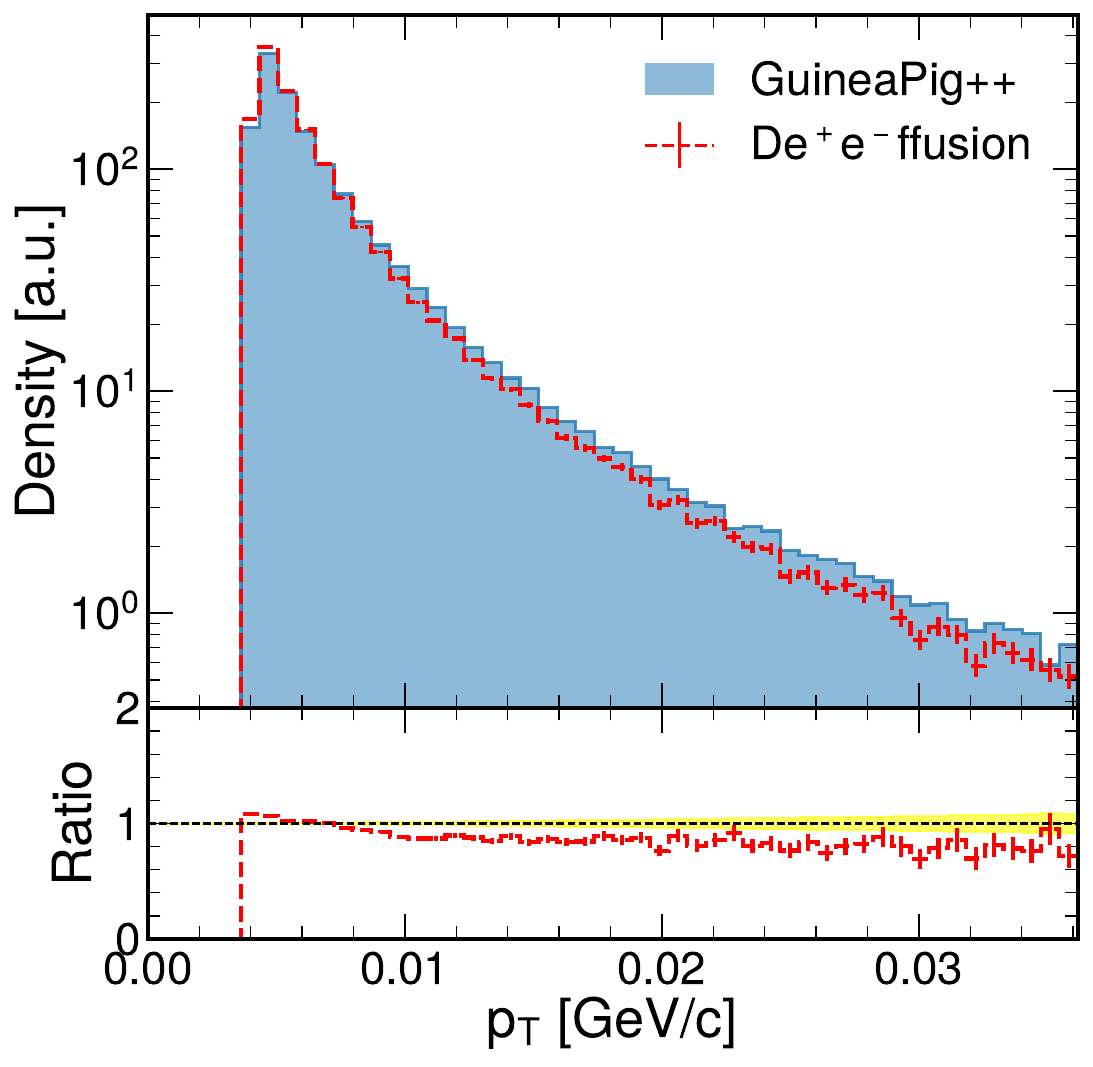}
\includegraphics[width=0.32\linewidth]{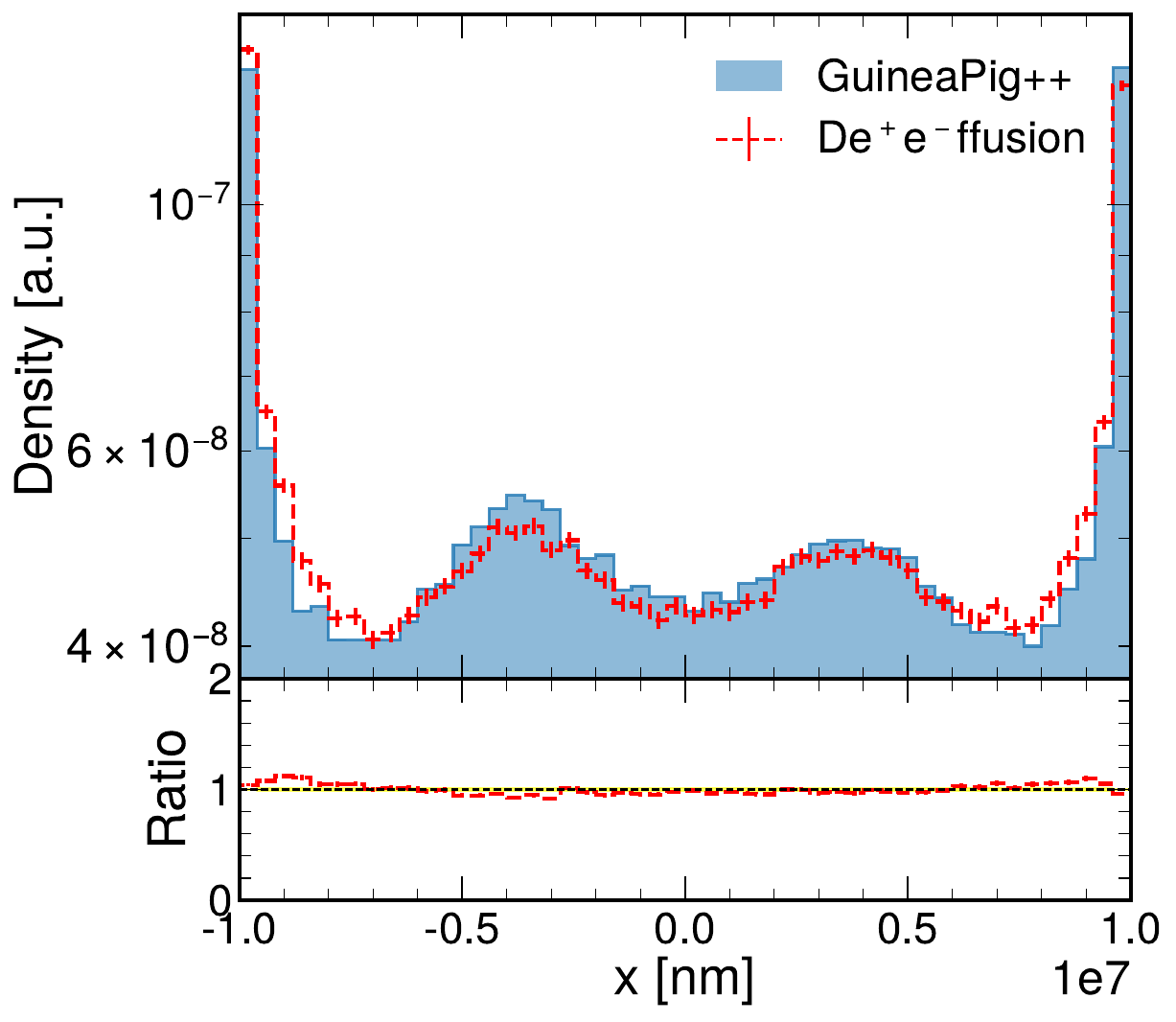}
\includegraphics[width=0.32\linewidth]{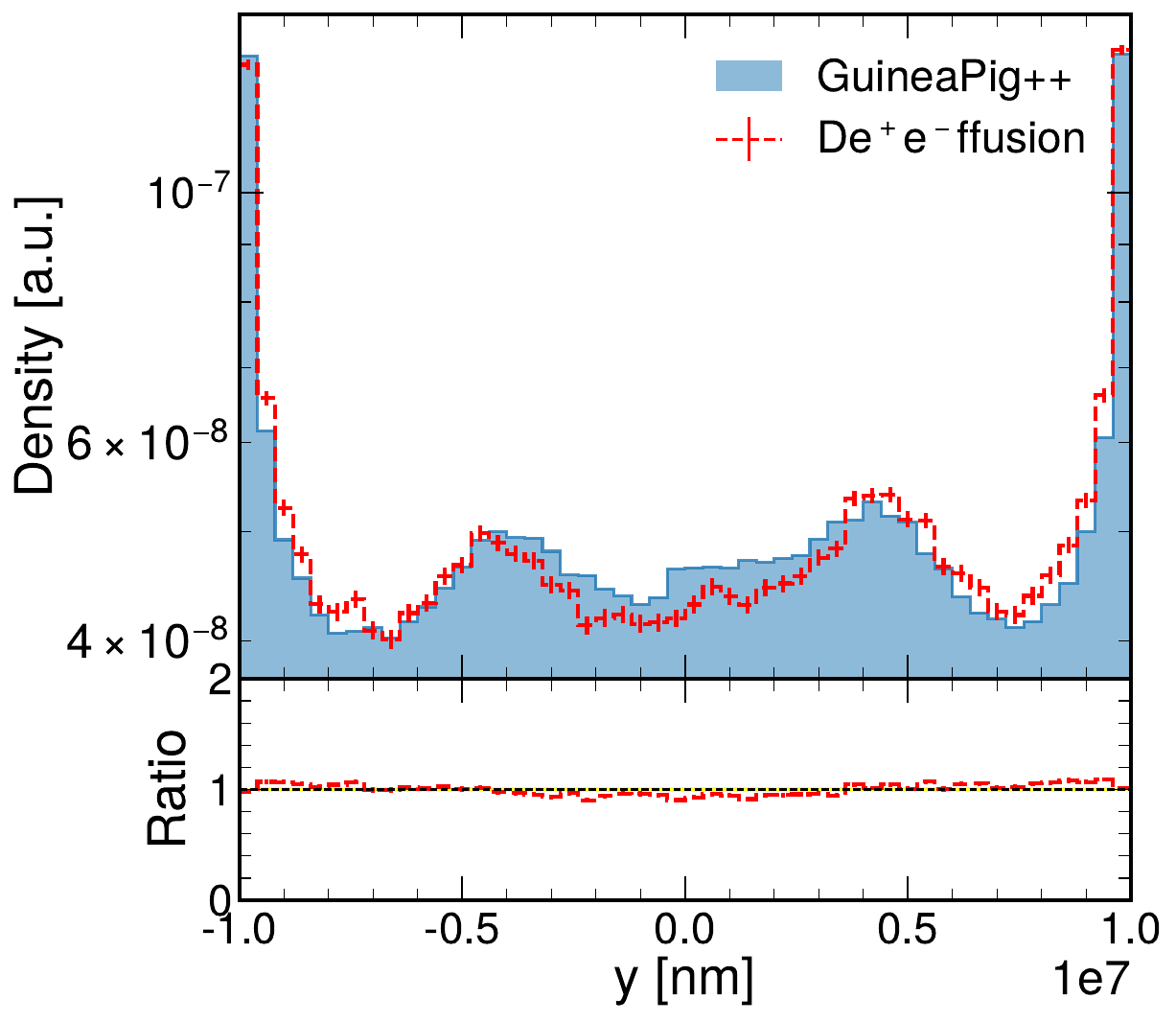}
\includegraphics[width=0.32\linewidth]{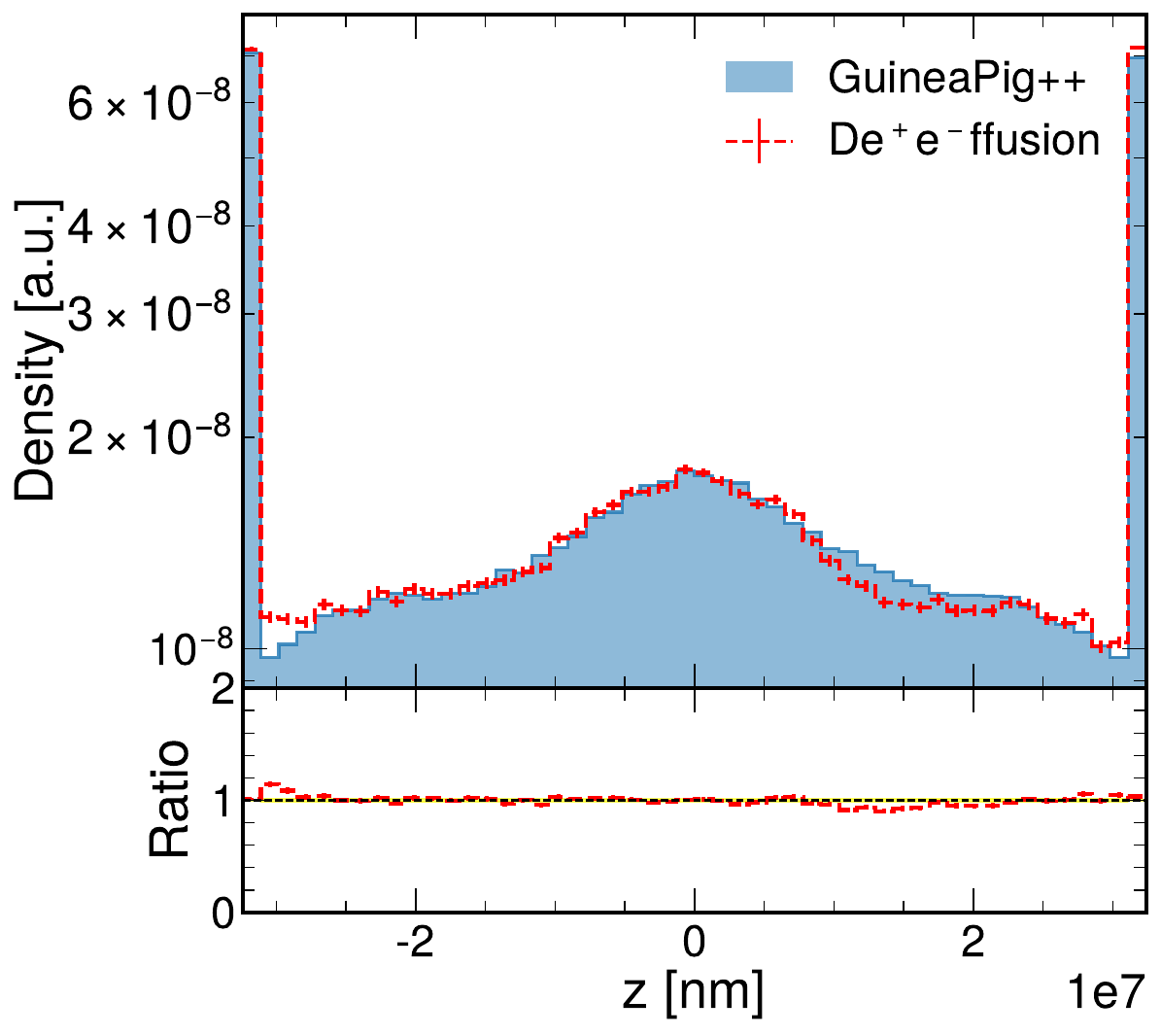}
\caption{\justifying Particle-level one-dimensional marginal distributions of $\beta_x$ (top-left), $\beta_y$ (top-middle), $\beta_z$ (top-right), $\abs{E}$ (center-left), $\abs{p}$ (center-middle), $\pT$ (center-right), $x$ (bottom-left), $y$ (bottom-middle), and $z$ (bottom-right), restricted to the subset of IPC
particles that satisfy the fiduciality condition of Eq.~\eqref{eq:reach} for the innermost CLD vertex-barrel layer. For each panel the filled blue histogram corresponds to {\gpig}, red dashed is the diffusion model {\deffusion}. All histograms are normalized to unit area. The error bars represent the statistical uncertainty in each bin. The bottom panels in each plot show the ratio of the {\deffusion} model's prediction of each observable to their truth {\gpig} values and the yellow-shaded region indicates the statistical uncertainty in the truth.}
\label{fig:marg_det_reach}
\end{figure*}

The marginal distributions of Fig.~\ref{fig:marg} characterize the full IPC final state, integrated over the entire seven-dimensional single-particle phase space. For detector-occupancy and radiation-damage studies, however, the
overwhelming majority of these particles are irrelevant since the IPC spectrum is both very soft and very forward. As such, the majority of pairs spiral tightly in the solenoidal field and curl back toward the beam pipe long
before reaching any active sensor. Only a small fraction of particles reach the fiducial volume of the detector, \ie are produced with sufficient transverse momenta and small enough polar angles to intersect at least the innermost detector layer and thus contribute to the occupancy. Therefore, in Fig.~\ref{fig:marg_det_reach} we restrict the comparison between {\gpig} and {\deffusion} to the subset of particles that satisfy the fiduciality condition for the innermost vertex-barrel layer (CLD detector~\cite{CLD:2019}).

The fiduciality criterion follows from considering the trajectory of a charged relativistic particle in a uniform axial magnetic field $\vec{B} = B_0\,\widehat{z}$ and is derived as follows. Starting from the relativistic Lorentz force equation, we have:
\begin{equation}
  \frac{d\vec{p}}{dt} = q\,(\vec{v}\times\vec{B}),
  \qquad
  \vec{p} = \gamma m \vec{v}
  \;\Longrightarrow\;
  \frac{d\vec{v}}{dt} = \vec{v}\times\vec{\omega}_c,
\end{equation}
with cyclotron frequency $\vec{\omega}_c = q\vec{B}/\gamma m$. Since the magnetic field does no work, the energy and hence the Lorentz factor $\gamma$ are conserved, and the motion decomposes into uniform translation along $\widehat z$ and circular motion in the transverse plane. Integrating the equations of motion with the production vertex approximated by the interaction point $(x_0,y_0,z_0)\simeq(0,0,0)$ yields a helix whose transverse radial extent is given by:
\begin{equation}
  r(z) = 2R\,\left|\sin\!\left(\frac{z\tan\theta}{2R}\right)\right|,
  \label{eq:helix}
\end{equation}
where $R = \pT/qB$ is the Larmor radius, $\pT$ the transverse momentum, and $\theta$ the polar angle of the particle momentum. Modeling the innermost vertex-barrel layer as a cylindrical shell of radius $r_\mathrm{det}$ and half-length $z_{\max}$, a particle of given $\pT$ first intersects the layer at the minimal polar angle for which the helix reaches $r = r_\mathrm{det}$ while still within the longitudinal acceptance $|z| \le z_{\max}$. Setting $r(z_{\max}) = r_\mathrm{det}$ in Eq.~\eqref{eq:helix} gives the fiducial boundary in the $\pT$-$\theta$ plane, as:
\begin{equation}
  \pT\,\left[\sin\!\left(\frac{\tan\theta}{\pT}\cdot
  \frac{qBz_{\max}}{2}\right)\right] = \frac{qBr_\mathrm{det}}{2}.
  \label{eq:reach}
\end{equation}
A particle is retained if and only if its $\pT$ and $\theta)$ make it fiducial given Eq.~\eqref{eq:reach}, \ie its maximum radial excursion carries it out to $r_\mathrm{det}$ within the barrel length. For the CLD vertex geometry, fewer than one in $\mathcal{O}(10^{3})$ IPC particles are retained. 

Figure~\ref{fig:marg_det_reach} repeats the one-dimensional marginal comparison of Fig.~\ref{fig:marg} after imposing the fiduciality condition~\eqref{eq:reach}. As expected, the surviving particles populate the high-$\pT$, large-$|\beta_{x,y}|$ tails of the parent distributions, because the selection removes the soft, tightly curling core that dominates the inclusive spectra and isolates precisely the population that drives sensor occupancy in the inner tracker. This is a considerably more stringent test of the surrogate than the inclusive marginals, since it probes sparsely populated tail regions of phase space where statistical fluctuations are larger and where any mismodeling of the momentum or angular tails are most visible. As such, while {\deffusion} reproduces the fiducial subset reasonably well across all observables, with the ratio panels remaining close to unity throughout with deviations on the order of statistical uncertainty, it is not at the percent-level agreement seen in Fig.~\ref{fig:marg} across the board. In particular, the high $E$, $\abs{p}$, and $\pT$ tails are under-predicted by approximately 10--15\%.

If desired it is straightforward to repeat the training by first removing the particles that are irrelevant for a given detector geometry to achieve a better agreement with the original {\gpig} simulation. On the other hand this limits the usability of the samples because the fiducial volume of the detector depends on the specific detector geometry. Our primary goal is to produce a detector agnostic simulation.

\subsection{Detector-Level Evaluation: Classifier Two-Sample Test}

The particle-level comparisons above demonstrate that {\deffusion} reproduces the one-dimensional kinematic, angular, and spatial marginals of {\gpig}. However, agreement in projected observables does not by itself guarantee that the full high-dimensional event structure is faithfully modeled. In particular, small mismodeling in correlations between position, momentum, charge, and process composition can be amplified after propagation through the detector material and magnetic field. Therefore, we perform a detector-level classifier two-sample test, which provides a more global and sensitive measure of agreement between the generated and reference distributions.

\begin{figure*}[t]
\centering
\includegraphics[width=0.49\linewidth]{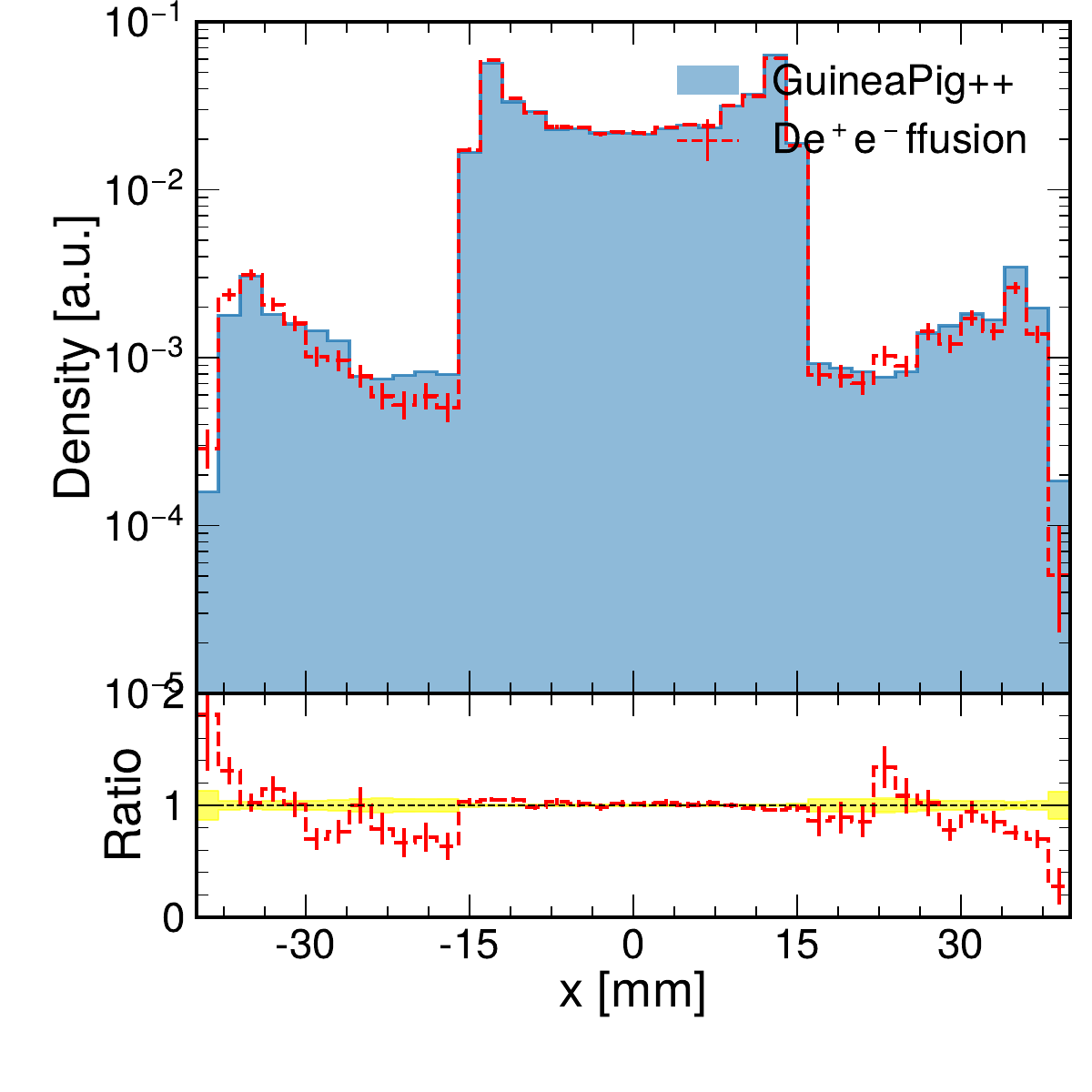}
\includegraphics[width=0.49\linewidth]{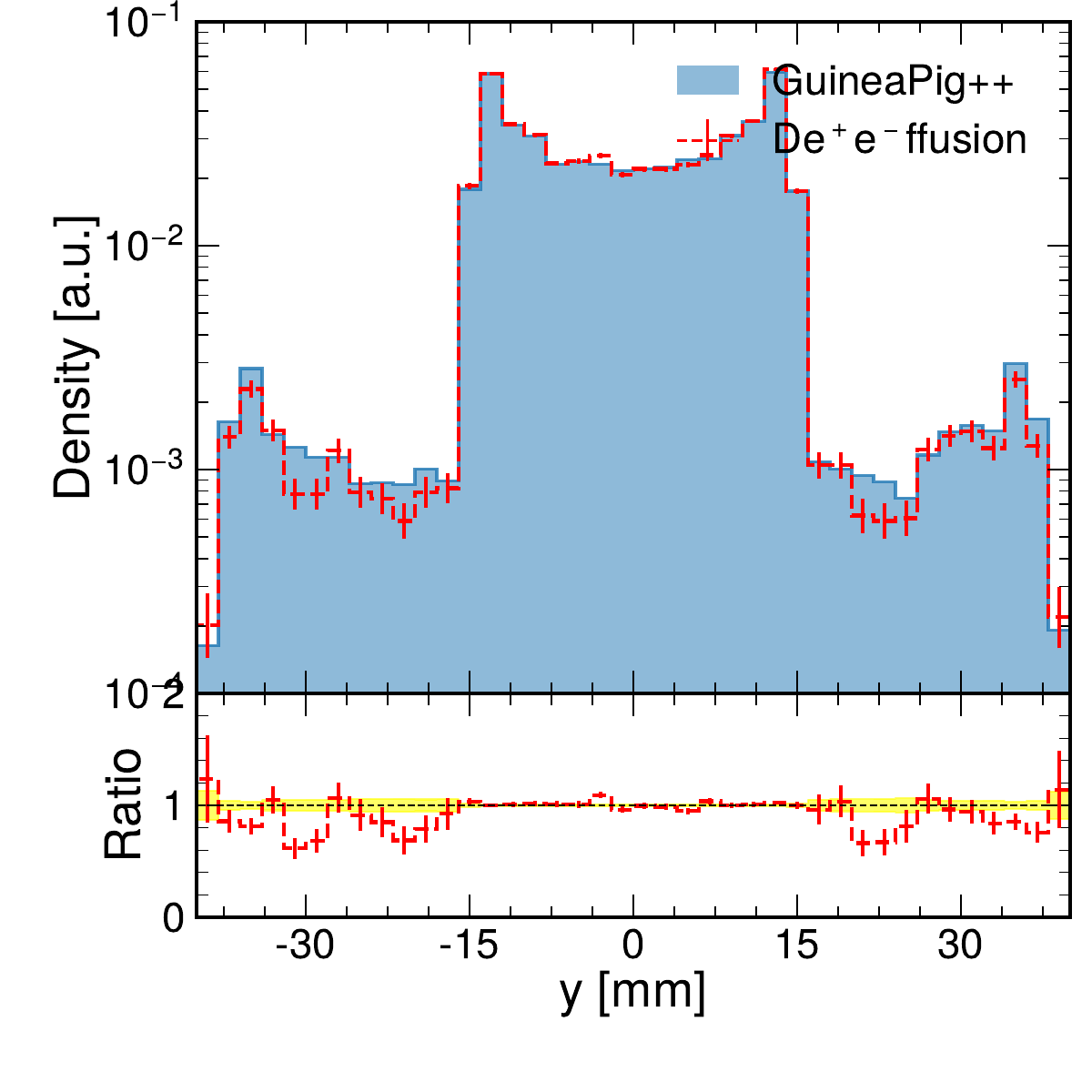}
\includegraphics[width=0.49\linewidth]{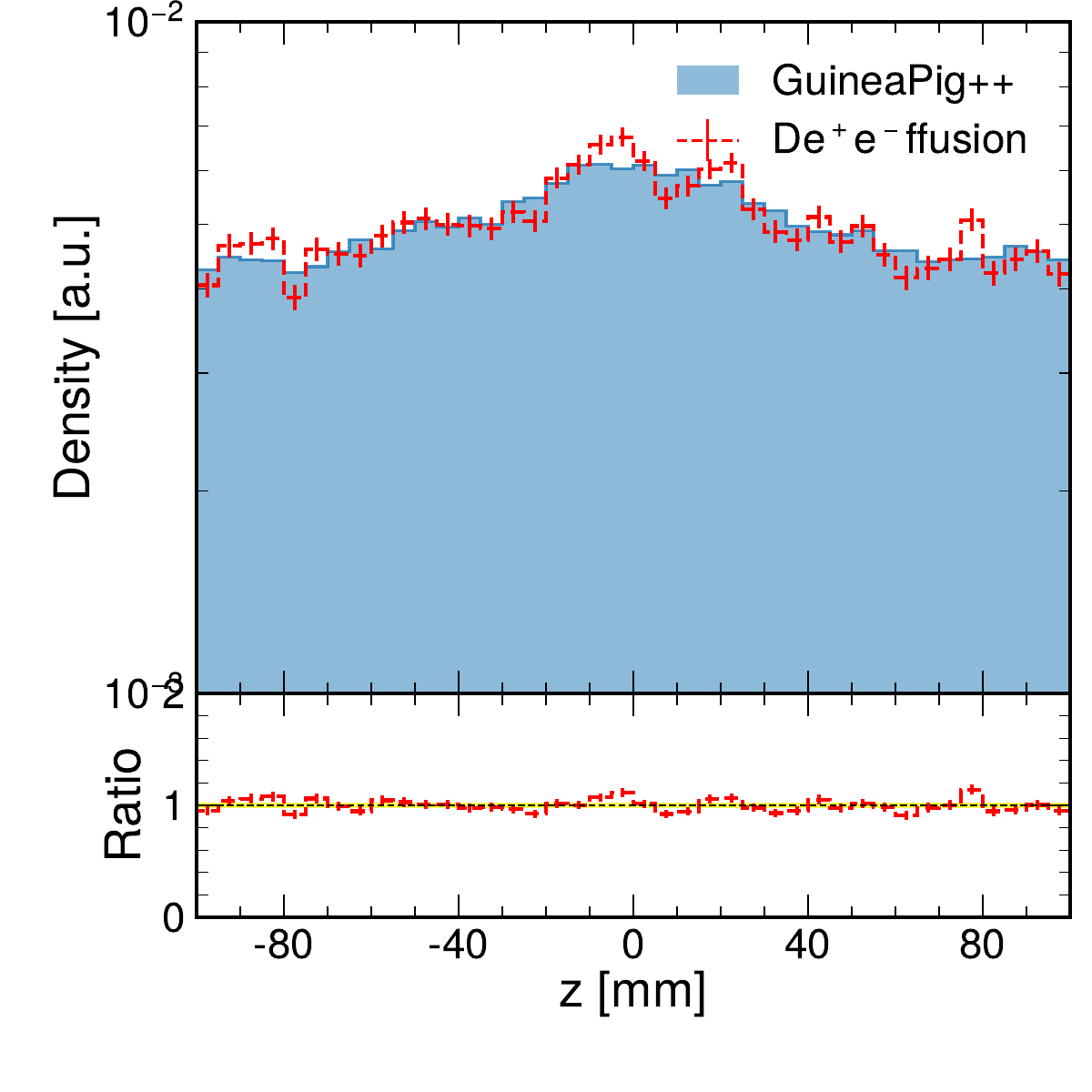}
\includegraphics[width=0.49\linewidth]{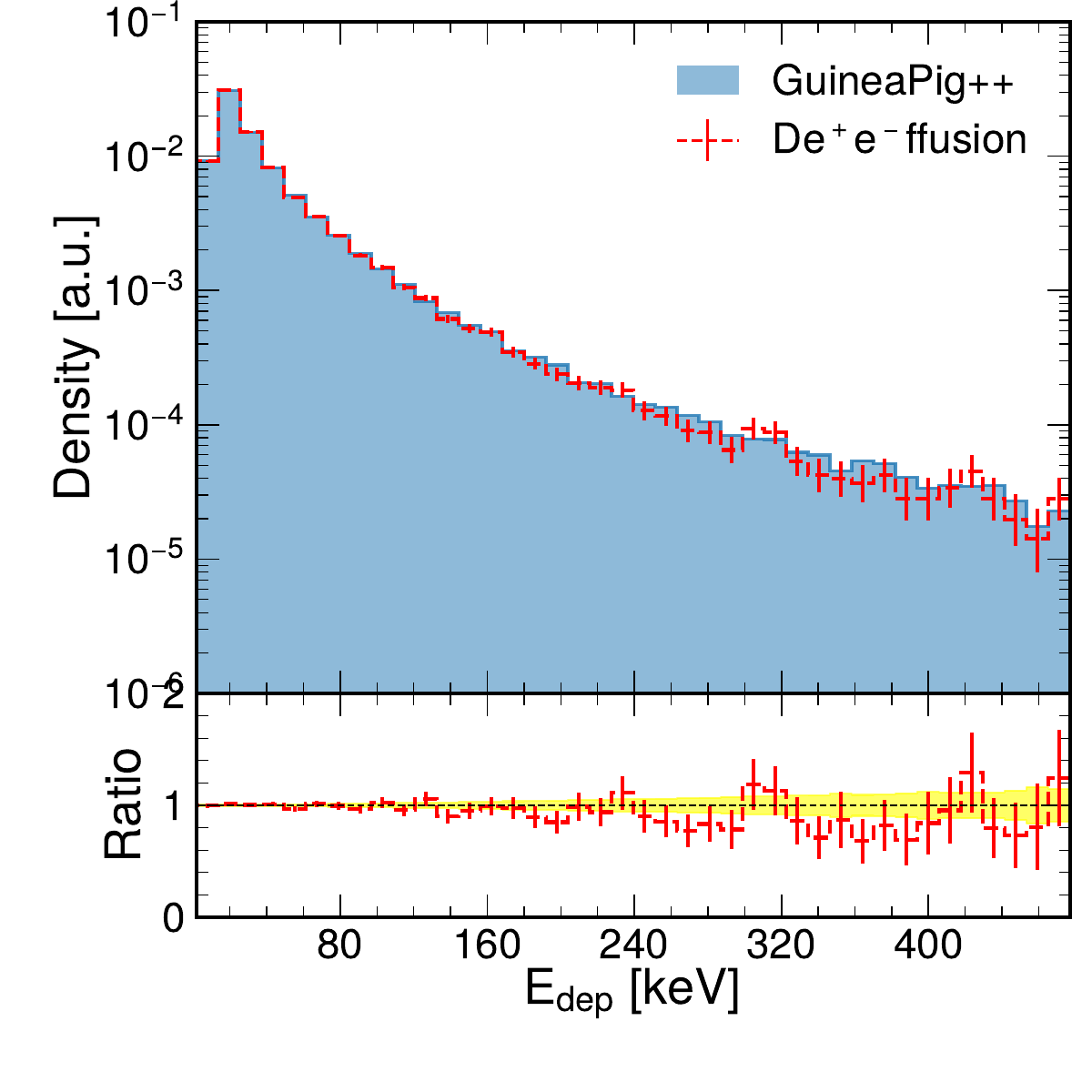}
\caption{\justifying Detector-level comparison of hits in the CLD vertex detector following {\gfour} simulation via \texttt{ddsim}. Shown are the $x$ (top left) $y$ (top right), $z$ (bottom left), and $E_\mathrm{dep}$ (bottom right) distribution for {\gpig} (solid blue) and {\deffusion} (red dashed). All histograms are normalized to unit area. The error bars represent the statistical uncertainty in each bin. The ratio of the {\deffusion} model's prediction to the truth {\gpig} value is shown in the bottom panel. The yellow-shaded region indicates the statistical uncertainty in the truth.}
\label{fig:cld}
\end{figure*}

Both {\gpig} and {\deffusion} events are propagated through a {\gfour}~\cite{Geant4:2003} simulation of the CLD~\cite{CLD:2019} vertex detector. In particular, generator-level background particles are propagated through the full CLD detector geometry from the \textsc{k4geo} compact geometry description using \texttt{ddsim}~\cite{Petri2017}, the DD4hep~\cite{Frank_2014} simulation driver from the \textsc{Key4Hep}~\cite{Ganis:2021vgv} software ecosystem. SimHits are subsequently stored in the EDM4hep data format~\cite{refId0}. 

The corresponding hit-level marginals are shown in Fig.~\ref{fig:cld}. The $x$ and $y$ distributions exhibit the expected structure from the cylindrical barrel geometry and the curling of low-$\pT$ particles in the solenoidal magnetic field. {\deffusion} reproduces the main geometric features of these distributions, including the sharp transitions associated with the active barrel acceptance and the lower-density tails at larger transverse displacement. The ratio panels remain close to unity across the bulk of the distributions, with the largest deviations appearing near detector edges and in low-density regions where the distributions change rapidly.

The longitudinal hit distribution, $z$, is also well reproduced over the full barrel extent. This is a particularly important validation because the detector-level $z$ profile is sensitive to the process-dependent generation structure, as well as to the subsequent transport of the low-momentum pairs through the detector. The agreement in the $z$ hit distribution therefore confirms that the process-conditioned diffusion model does not merely reproduce the particle-level longitudinal marginals, but also propagates them into the correct detector occupancy pattern.

The test is performed on the resulting detector response in the three innermost vertex barrel layers, using the hit-level collection
$
\{(r,\phi,z,\Delta t,\Delta E)\}_h
$
as the input particle point cloud. This choice makes the test more stringent than the particle-level comparisons since the classifier is not only sensitive to the primary IPC phase-space distribution, but also to the induced detector occupancy pattern, the spatial distribution of deposited energy, and any correlations that may get amplified through the magnetic field and detector material.

We train a transformer-based classifier $h_\phi : \mathcal{H} \to [0,1]$ to distinguish {\gpig} events from {\deffusion} events, assigning labels $y=0$ to the reference sample and $y=1$ to the generated sample. The classifier operates on unordered hit collections, making it naturally suited to the variable-size, high-dimensional, set-valued structure of the detector response. In this setting, the classifier acts as a proxy for the likelihood-ratio test between the two detector-level distributions. If the generated and reference samples were perfectly identical, no classifier could perform better than random guessing, and the expected area under the ROC curve would be $\mathrm{AUC}=0.5$. Conversely, an AUC significantly above $0.5$ would indicate that the classifier has found residual information that separates the two samples, corresponding to detector-level mismodeling.

The classifier $h_\phi$ is implemented as a permutation-invariant transformer encoder operating on hit point clouds. Each hit $(r,\phi,z,\Delta t,\Delta E)_h$ is first projected to a $d_\mathrm{model}=128$ embedding through a learned linear layer; no ad-hoc positional encoding is added, so the network is equivariant under permutations of the hit ordering. The embeddings are then processed by a stack of $N_\mathrm{layers}=4$ standard transformer encoder blocks~\cite{Vaswani:2017}, each with $h=8$ self-attention heads (with per-head dimension $d_k = d_\mathrm{model}/h = 16$), feed-forward width $d_\mathrm{ff}=128$, GELU activations, pre-norm LayerNorm, and dropout $p=0.1$. A key-padding mask is propagated through all attention layers so that the zero-padded slots in each batch do not contribute to either the attention logits or the readout. A mask-aware mean-pooling over the hit dimension produces a single event embedding, which is passed to a two-layer MLP head ($d_\mathrm{model}\to d_\mathrm{model}\to 1$) with LayerNorm, GELU, and dropout, returning the classifier logit. The resulting model has approximately 400k trainable parameters.

Samples are split 80/10/10 into training, validation, and test sets. The network is trained with the binary cross-entropy loss on the logits, with optional inverse-frequency class reweighting to mitigate any residual class imbalance. As with the training of the main DDPM, we use the {\textsc{AdamW}}~\cite{Loshchilov:2017adamw} optimizer with base learning rate $\eta_0 = 2\times 10^{-4}$, weight decay $\lambda = 10^{-2}$, $(\beta_1,\beta_2)=(0.9,0.999)$, and gradient-norm clipping at $1.0$. The learning rate follows a cosine decay schedule from $\eta_0$ to $\eta_{\min}=10^{-6}$ over $30$ training epochs at a batch size of $64$. The checkpoint with the highest validation AUC is retained for the final test, and the AUC and ROC curve quoted below are evaluated on the held-out test split.

On the held-out test set, the classifier achieves $\mathrm{AUC} = 0.553 \pm 0.016$, where the uncertainty reflects the standard deviation across five independent training runs with different random seeds controlling both the train-validation split and the weight initialization, with individual runs spanning $[0.540,\,0.576]$. As we will see later in this section, this value is only modestly above the random-classification limit, indicating that the two detector-level distributions are close but not perfectly indistinguishable. Importantly, the classifier is trained directly on the high-dimensional hit clouds, so this result constrains not only the one-dimensional hit marginals, but also the correlations among hit position, deposited energy, and event occupancy. The fact that the classifier is unable to reach a large separation power therefore provides evidence that {\deffusion} preserves the detector-relevant structure of the IPC background to good accuracy.

To place this result on a firmer statistical footing we perform two complementary experiments. First, we draw 50 independent instances of the classifier $h_\phi$ with randomly sampled weights and no training, evaluate each on the same held-out validation split, and record the resulting AUC. Across these 50 random instantiations we obtain $\mathrm{AUC}_{\mathrm{rand}} = 0.519\pm0.014$, with individual runs spanning the interval $[0.501,\, 0.552]$. This confirms that an untrained network produces scores consistent with a uniform distribution, as expected, and establishes $\mathrm{AUC}=0.5$ as the empirical chance level for this architecture and dataset.

\begin{figure}
    \centering
\includegraphics[width=\linewidth]{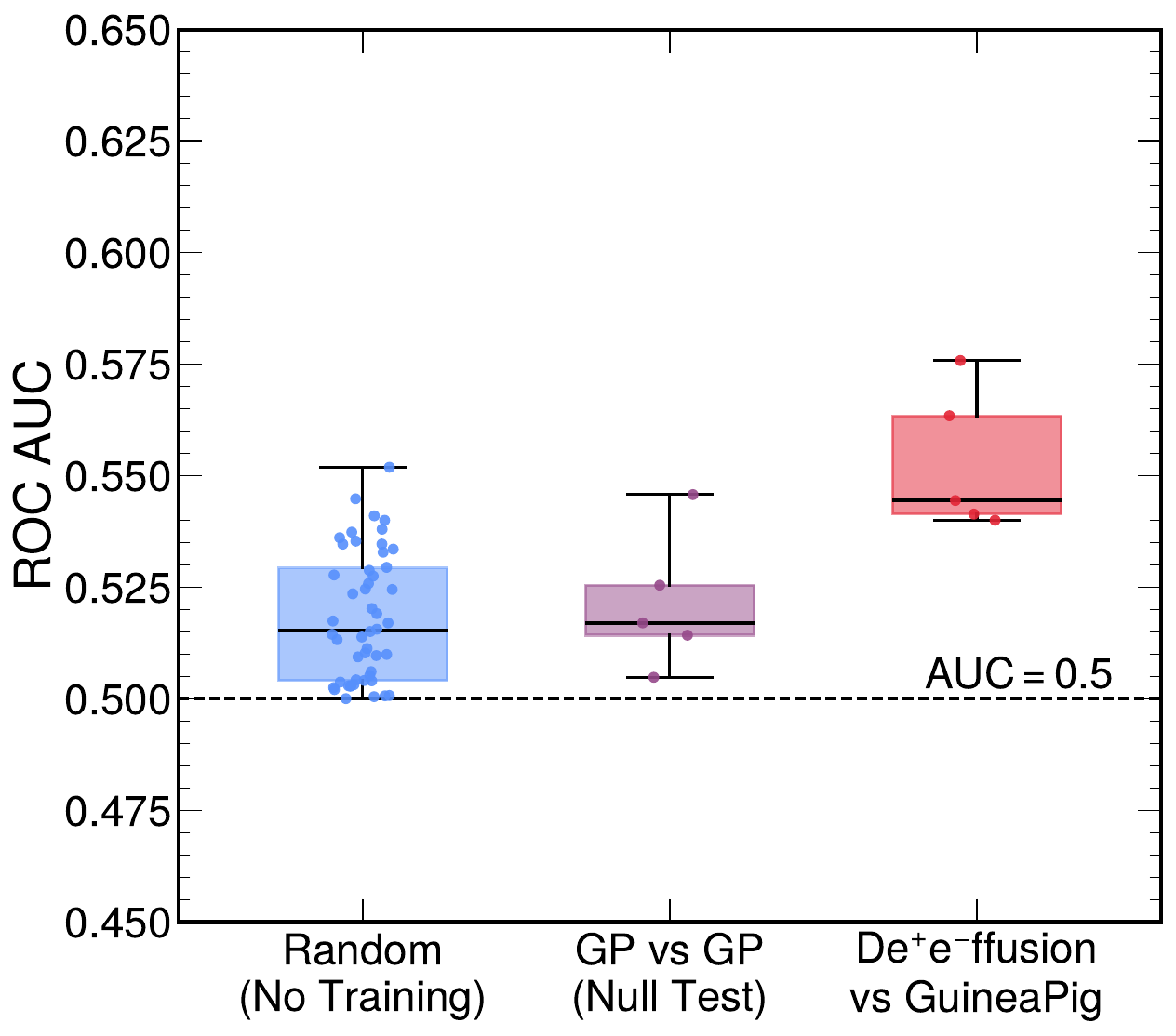}
    \caption{\justifying Box-and-whisker plot showing the distribution of the classifier ROC $\max\{\rm{AUC},1-\rm{AUC} \}$ across independent runs for the Random Classifier (blue left), {\gpig} vs {\gpig} null test (purple middle), and {\gpig} vs {\deffusion} (red right) test. Each box spans the inter-quartile range with the median shown as a horizontal line, the whiskers extend to the full range, and individual runs are overlaid as points.}
    \label{fig:auc_box}
\end{figure}

Second, we characterize the finite-sample noise floor of a trained classifier; the test is motivated by the consideration that due to the relatively small sample sizes considered in this study, two samples drawn from the same distribution may exhibit statistical fluctuations that a trained classifier may be able to distinguish. To this end, we replace the two-sample problem with a null experiment in which both classes are drawn from the same {\gpig} sample. We randomly partition the {\gpig} sample into two equal halves, assign class labels $y=0$ and $y=1$ arbitrarily, and train the classifier following the identical procedure described above. This experiment is repeated for five independent random partitions. The trained classifiers achieve $\mathrm{AUC}_{\mathrm{null}} = 0.522\pm 0.024$, with individual runs spanning $[0.505,\,0.546]$. The fact that a fully trained classifier operating on two statistically identical samples cannot surpass $\mathrm{AUC}\approx 0.51$--$0.55$ quantifies the irreducible noise floor of this test at the available sample size. As such, we stress that any AUC attributable to genuine distributional differences between {\deffusion} and {\gpig} must be interpreted against this baseline rather than the idealized $\mathrm{AUC}=0.5$ limit, as summarized in Fig.~\ref{fig:auc_box}. The primary result, $\mathrm{AUC}=0.553\pm 0.016$, lies within the upper tail of the null distribution and is only modestly separated from it. The seed-to-seed standard deviation of $0.016$ is comparable to the width of the null distribution ($\sigma=0.024$), confirming that the result is stable across realizations.

Finally, the deposited-energy spectrum, $E_\mathrm{dep}$, spans several orders of magnitude and provides a complementary test of the detector response. The generated sample follows the {\gpig} reference closely across the steeply falling spectrum, including the high-$E_\mathrm{dep}$ tail where statistics are more limited. Since $E_\mathrm{dep}$ depends on the interaction of particles with detector material, this agreement suggests that the generated particles have sufficiently accurate momenta, incident angles, and spatial distributions to reproduce the main features of the simulated energy deposition.

Taken together, the classifier AUC and the hit-level marginal comparisons provide a detector-level validation of {\deffusion}. The residual separation power of the classifier indicates that small differences between the generated and reference samples remain, as expected for a fast generative surrogate. However, the low AUC, combined with the near-closure of the $x$, $y$, $z$, and $E_\mathrm{dep}$ distributions, shows that these differences are subleading at the level of the CLD vertex-barrel response. This supports the use of {\deffusion} as a high-fidelity fast-simulation surrogate for IPC-induced detector occupancies and beam-background studies at FCC-ee.

\subsection{Runtime}

\begin{figure}[t]
\centering
\includegraphics[width=\linewidth]{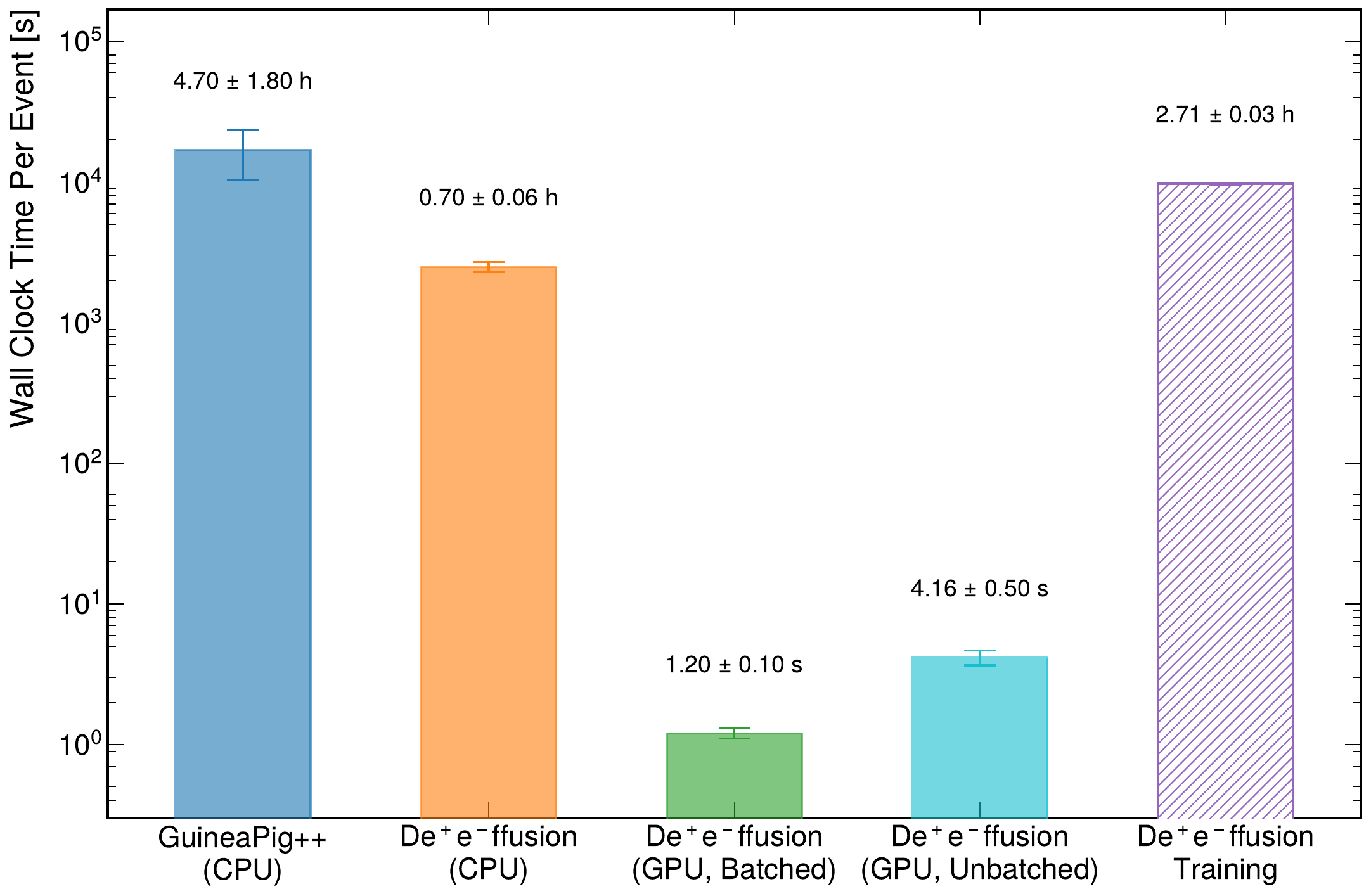}
\caption{\justifying Per-event generation wall-clock time for {\gpig} (left), {\deffusion} with unbatched CPU sampling (middle-left), batched GPU sampling (middle-right), and unbatched GPU sampling (right). The wall-clock time to train a {\deffusion} model is also shown for comparison. The error bars represent a one standard deviation interval over 3000 samples.}
\label{fig:runtime}
\end{figure}

We benchmark the per-event generation cost of the surrogate against the {\gpig} reference, which requires about 15\,000~seconds per event on a single AMD Milan 7713 CPU core, as shown in Fig.~\ref{fig:runtime}. To isolate the speedup due to the surrogate implementation itself, we first run {\deffusion} on the same hardware. We sample without batching (batch size of 1) on a single AMD Milan 7713 CPU core, a sub-optimal worst-case scenario, {\deffusion} produces a single event in about 2500~seconds, already a factor of 5 speedup over {\gpig}. The GPU support of {\deffusion} improves this substantially. On a single NVIDIA A100 (40\,GB) GPU with $T'=10^3$ steps and a sampling batch of 16 (1) events, {\deffusion} takes about 1.20 (4.16) seconds per event, a speedup of nearly 4 orders of magnitude over the reference. Subsampling the reverse trajectory to, for instance, $T'=250$ steps yields a further $4\times$ speedup at the cost of a small but quantifiable degradation in the tails of the distributions reported above.

Training the model takes about 450~seconds (close to 8~minutes) per epoch, amounting to about 9000~seconds until convergence over 20 epochs as shown in the final bar in Fig.~\ref{fig:runtime}. Training though is only a one-time exercise that does not need to be repeated while sampling. It is also worth noting that training the model from scratch still takes substantially less time than simulating a single bunch crossing in {\gpig}.

\section{Discussion and Outlook}
\label{sec:discussion}

In this work, we have introduced {\deffusion}, a process-conditioned denoising diffusion probabilistic model fast simulation surrogate for the IPC component of FCC-ee beam-induced backgrounds. The model achieves approximately percent-level fidelity on every one-dimensional particle-level marginal we have examined, including the challenging longitudinal position distribution, and is statistically consistent with {\gpig} at the detector level under a classifier two-sample test, all at $\mathcal{O}(10^3)$ wall-clock speedup. Several modeling choices proved essential. First, working in the unsquashed velocity coordinate $\bu=\mathrm{atanh}(\bbeta)$ removes the lightlike boundary from the diffusion domain. Second, treating the QED subprocess as a conditioning input, which is never noised, makes the $z$ distribution reproducible; the unconditional model, on the other hand, averages the three channels and loses the multi-modal nature entirely. Third, the inverse-frequency reweighting of the score-matching loss prevents the rare BW and BH channels from being washed out by the much more numerous LL particles.

The long-term vision of {\deffusion} is to solve the computational challenges encountered by the FCC-ee collaborations in performing beam background studies that rely on high-statistics samples and to provide a fast simulation framework for both experimentalists and phenomenologists. At present, the simulation of $e^+e^-$ incoherent pairs is often an $\mathcal{O}(1)$ fraction of all computational expense in an FCC-ee beam-background study. As such, if we can accelerate it by nearly four orders of magnitude as demonstrated in this study, in addition to making it compatible with hardware accelerators (e.g., GPUs), the IPC simulation would no longer be a bottleneck for data analysis.

\begin{figure*}[htbp]
  \centering
  \includegraphics[width=\textwidth]{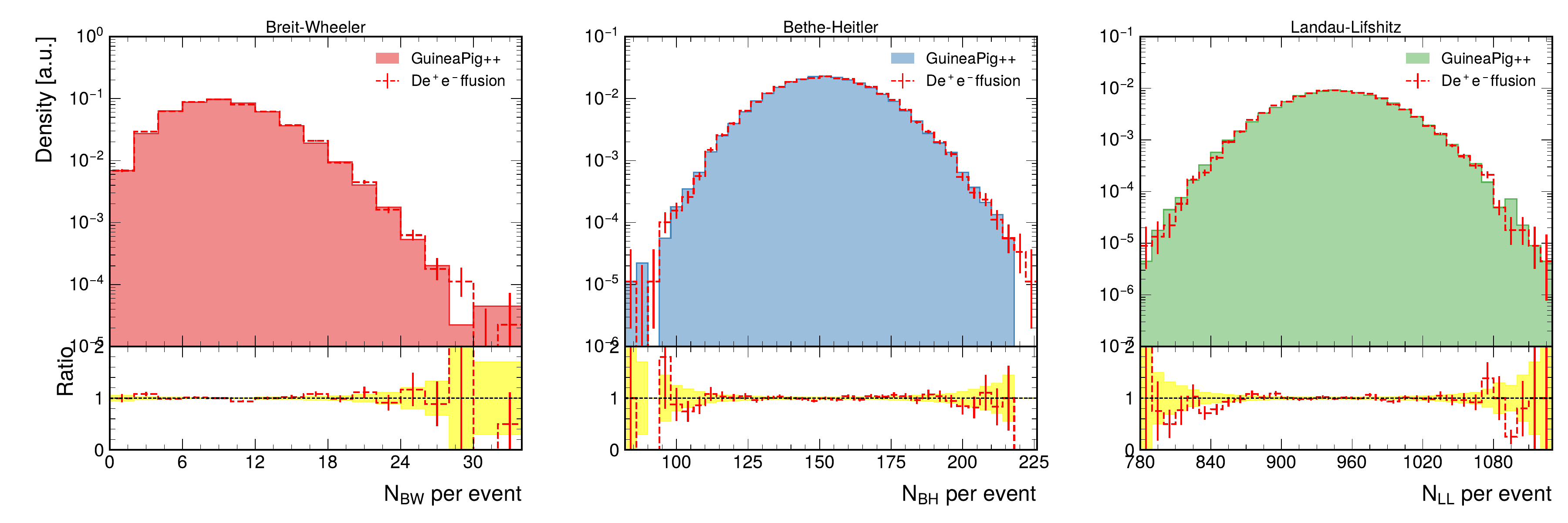}
  \caption{\justifying Particle-level per-event multiplicity split by QED subprocess: 
  Breit--Wheeler (left), Bethe--Heitler (center), and Landau--Lifshitz 
  (right), for \gpig (filled) and \deffusion (red dashed). All 
  histograms are normalized to unit area. The error bars represent the 
  statistical uncertainty in each bin. The bottom panels show the ratio 
  of the {\deffusion} prediction to the truth {\gpig} value; the 
  yellow-shaded region indicates the statistical uncertainty in the 
  truth.}
  \label{fig:mult_perprocess}
\end{figure*}

\begin{figure}
    \centering
    \includegraphics[width=\linewidth]{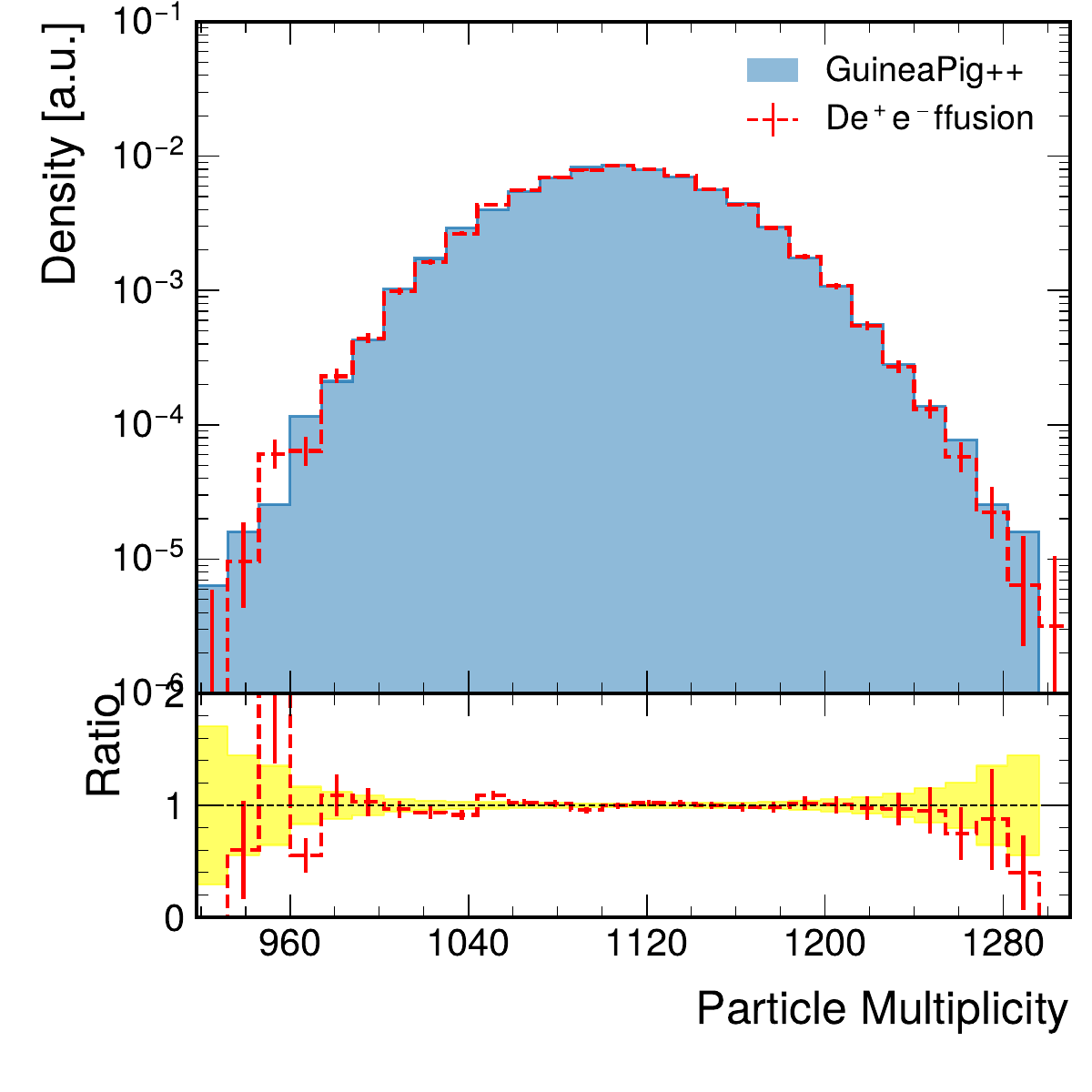}
    \caption{\justifying Total per-event particle-level multiplicity for {\gpig} 
  (filled blue) and {\deffusion} (red dashed). All histograms are 
  normalized to unit area. The error bars represent the statistical 
  uncertainty in each bin. The bottom panel shows the ratio of the 
  {\deffusion} prediction to the truth {\gpig} value; the 
  yellow-shaded region indicates the statistical uncertainty in the 
  truth.}
    \label{fig:mult_total}
\end{figure}

While the work presented in this manuscript represents a substantial advancement toward this goal, several crucial steps still remain. The first natural, and arguably most pressing, extension of the present work would be to condition on machine parameters, e.g., bunch sizes, intensities, and crossing angle, as inputs to the model to enable rapid scans of the parameter space during the FCC-ee design phase. This is particularly important because the method in its current formulation cannot extrapolate to unseen machine configurations. We emphasize, however, that this is a technical and not conceptual addition and training over multiple parameter sets would naturally mitigate this parameter dependence and is planned for future work. 

Additional steps also include extending the diffusion process to beamstrahlung photons, which would close the BIB loop at the level of the full bunch crossing. One could also replace the empirical multiplicity and composition draws by a learned latent prior, in the spirit of the cardinality head of \textsc{Parnassus}~\cite{Dreyer:2024parnassus}. We expect each of these to be a minor addition rather than a major leap.

It is also worth noting that, in parallel with machine-learning-side developments, the landscape of first-principles beam-beam simulation is itself rapidly evolving. In particular, \textsc{WarpX}~\cite{WarpX}, a massively parallel, GPU-accelerated particle-in-cell framework, is being actively developed as a modern alternative to {\gpig} for FCC-ee beam-beam studies, and reports performance improvements comparable to those of {\deffusion} on ILC simulation~\cite{kicsiny_fcc2026}. The relevant strong-field processes, such as beamstrahlung, radiative and elastic Bhabha scattering, and incoherent pair creation, have already been implemented, alongside FCC-ee-specific features such as the crossing angle with Lorentz-boosted frame and the crab-waist scheme. While these capabilities are still under active development and validation, \textsc{WarpX} would be a natural point of comparison for future work.

Finally, for widespread adoption and use, we envision that {\deffusion} would require a thoroughly documented, user-friendly interface comparable with that of {\gpig} and \textsc{WarpX}, for instance. This can be accomplished straightforwardly by exporting the model using the {\textsc{Onnx}} library~\cite{onnxruntime}, and wrapping it into a portable \texttt{pip}-installable package. This also enables the method to be bundled with commonly used FCC-ee software ecosystems such as \textsc{Key4Hep}, including interfaces to \texttt{ddsim} etc.~to enable rapid development.

The progress presented in this study paves the way for {\deffusion} as a fast, high-fidelity surrogate for simulating the beam-induced backgrounds for the FCC-ee. 

\section*{Data and Code Availability}

All code that implements the algorithms presented in Section~\ref{sec:method} is available at \url{https://github.com/umarsqureshi/Deeffusion}. The simulated datasets are available from the corresponding authors upon reasonable request.

\section*{Acknowledgments}
This work made use of resources provided by subMIT facility at the MIT Physics Department~\cite{Acosta:2025kgk} in addition to SLAC Shared Science Data Facility (S3DF) at SLAC National Accelerator Laboratory.  SLAC is operated by Stanford University for the U.S. Department of Energy's Office of Science. The work of the authors is supported by the U.S. Department of Energy under contract DE-AC02-76SF00515.
BM acknowledges the support of Schmidt Sciences. 
LG is supported by the DOE, Office of Science, Office of High Energy Physics under Award No. DE-SC0010010 and the Early Career Research program under Award No. DE-SC0026288.
The work of DN, USQ, and CV is sponsored by the U.S. Department of Energy, Office of Science under Contract No. DE-AC02-76SF00515.

\appendix
\section{Multiplicity and Event Composition}
\label{app:multiplicity}

\begin{figure}
    \centering
    \includegraphics[width=\linewidth]{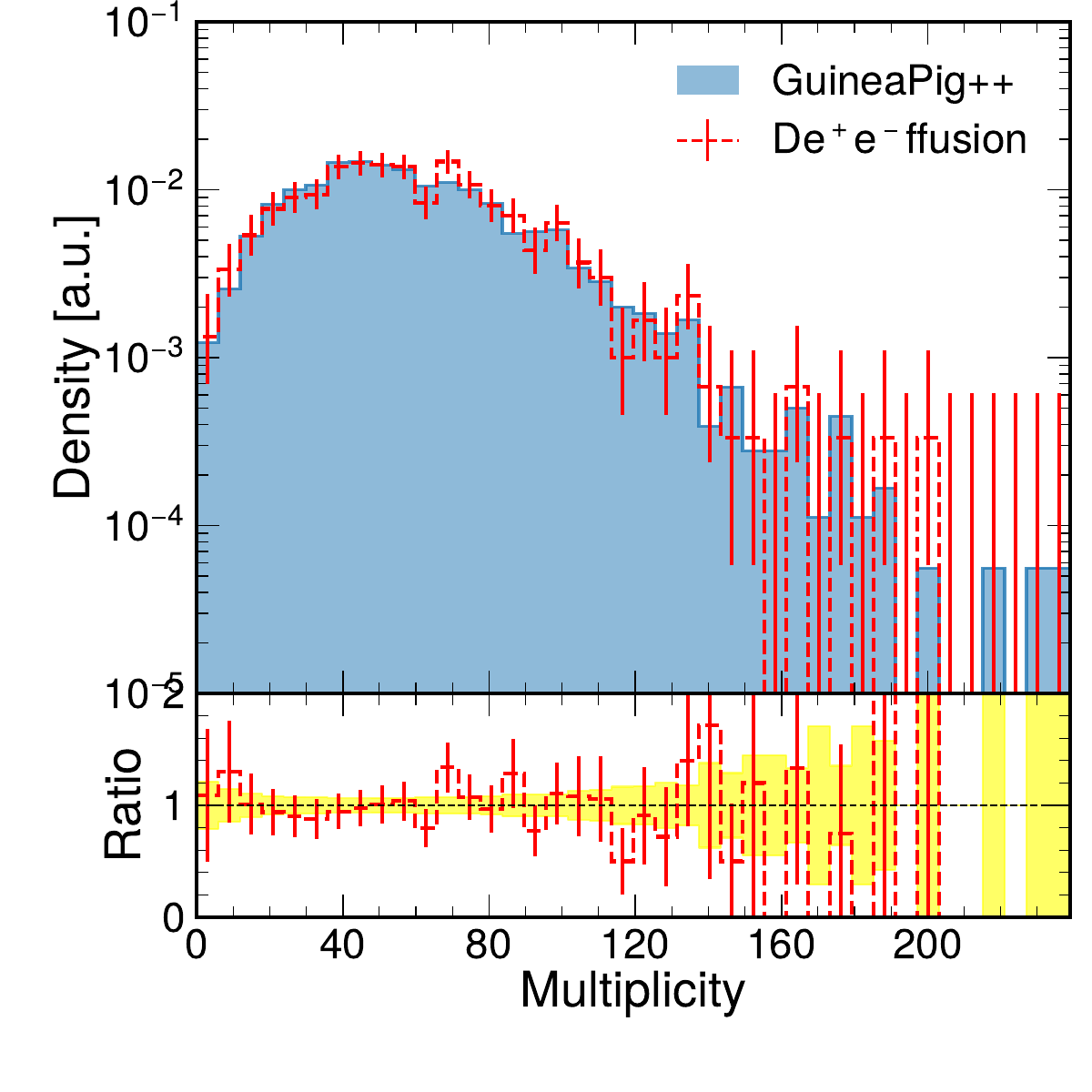}
    \caption{\justifying Per-event hit multiplicity in the CLD vertex-barrel layers after \texttt{ddsim} simulation, for \gpig (filled blue) and \deffusion (red dashed). All histograms are normalized to unit area. The error bars represent the statistical uncertainty in each bin. The bottom panel shows the ratio of the \deffusion prediction to the truth \gpig ground truth value; the yellow-shaded region indicates the statistical uncertainty in the truth.}
    \label{fig:mult_hits}
\end{figure}

As described in Section~\ref{sec:sampling}, {\deffusion}  does not learn the multiplicity distribution $p(K)$ directly. Instead, the empirical per-event multiplicity histogram $\{K_n\}_{n=1}^{N_\mathrm{ev}}$ and the per-event process composition $\mathbf{f}_n = (f^{\mathrm{BW}}_n, f^{\mathrm{BH}}_n, f^{\mathrm{LL}}_n)$ are cached from the training sample and supplied non-parametrically at sampling time. This handling is exact by construction up to the resolution of the empirical sample. In this appendix we verify that the sampling procedure accurately reproduces both the total and per-process multiplicity distributions, and that no observable bias is introduced.

Figure~\ref{fig:mult_total} compares the total per-event particle multiplicity for {\gpig} and {\deffusion}. The generated distribution tracks the reference closely across the full range, with the ratio panel remaining flat at unity throughout the bulk and only mild deviations in the statistically sparse high and low-multiplicity edges. This confirms that the non-parametric multiplicity draw and the subsequent masking of padded slots introduce no measurable bias in the overall event size.

A more stringent test is the multiplicity decomposed by QED subprocess, which probes whether the cached composition $\mathbf{f}_n$ together with the process-conditioned generation correctly reproduces the number of particles attributed to each channel. Figure~\ref{fig:mult_perprocess} shows the per-event counts $N_\mathrm{BW}$, $N_\mathrm{BH}$, and $N_\mathrm{LL}$ for the three production mechanisms. The mean multiplicities differ by roughly two orders of magnitude across channels, reflecting the relative cross sections $\sigma_\mathrm{BW}:\sigma_\mathrm{BH}:\sigma_\mathrm{LL} \approx 1:14:85$. {\deffusion} reproduces all three distributions within their statistical uncertainties, with the ratio panels close to unity across the bulk of each spectrum. 

Finally, we verify that the correct multiplicity is preserved after propagation through the detector. It is conceivable that even if the particle-level multiplicity is well-modeled, subtle mismodeling in the kinematics may result in differences in detector-level hit multiplicity, owing to a different number of particles making it to the detector. Figure~\ref{fig:mult_hits} shows the per-event hit multiplicity in CLD vertex-barrel layers, obtained by passing both {\gpig} and {\deffusion} events through the \textsc{Geant4} simulation with \texttt{ddsim}. The hit count is substantially lower than the particle-level multiplicity of Fig.~\ref{fig:mult_total}, since most low-$\pT$ pairs curl in the solenoidal field and never reach the active layers. {\deffusion} reproduces the hit-multiplicity distribution well, with the ratio remaining close to unity across the bulk and deviations only in the sparsely populated tails. 

\bibliography{deeffusion}
\end{document}